\documentclass[aps,pre,twocolumn,floatfix]{revtex4}

\usepackage{bm}
\usepackage{graphicx}
\usepackage{amssymb,amsfonts,amsmath}
\usepackage{epsf}
\usepackage{color}
\usepackage{hyperref}
\usepackage{subfigure}
\usepackage{epstopdf}
\usepackage{mathrsfs}

\usepackage{upgreek}

\newcommand{\be}{\begin{equation}}
\newcommand{\ee}{\end{equation}}
\newcommand{\bea}{\begin{eqnarray}}
\newcommand{\eea}{\end{eqnarray}}

\begin{document}

\title{\bf Quantum local-equilibrium approach to dissipative hydrodynamics}

\author{Jo\"el Mabillard}
\email{Joel.Mabillard@ulb.be; \vfill\break ORCID: 0000-0001-6810-3709.}
\author{Pierre Gaspard}
\email{Gaspard.Pierre@ulb.be; \vfill\break ORCID: 0000-0003-3804-2110.}
\affiliation{Center for Nonlinear Phenomena and Complex Systems, Universit{\'e} Libre de Bruxelles (U.L.B.), Code Postal 231, Campus Plaine, B-1050 Brussels, Belgium}

\vskip 0.5 cm

\begin{abstract}
The macroscopic hydrodynamic equations are derived for many-body systems in the local-equilibrium approach, using the Schr\"odinger picture of quantum mechanics.  In this approach, statistical operators are defined in terms of microscopic densities associated with the fundamentally conserved quantities and other slow modes possibly emerging from continuous symmetry breaking, as well as macrofields conjugated to these densities.  Functional identities can be deduced, allowing us to identify the reversible and dissipative parts of the mean current densities, to obtain general equations for the time evolution of the conjugate macrofields, and to establish the relationship to projection-operator methods.  The entropy production is shown to be nonnegative by applying the Peierls-Bogoliubov inequality to a quantum integral fluctuation theorem. Using the expansion in the gradients of the conjugate macrofields, the transport coefficients are given by Green-Kubo formulas and the entropy production rate can be expressed in terms of quantum Einstein-Helfand formulas, implying its nonnegativity in agreement with the second law of thermodynamics.  The results apply to multicomponent fluids and can be extended to  condensed matter phases with broken continuous symmetries.
\end{abstract}

\maketitle

\section{Introduction}

In fluids, the transport processes due to viscosity, heat conduction, and diffusion dissipate energy and produce entropy, which generates irreversibility at the macroscale.  Importantly, these dissipative effects should be taken into account in the equations of hydrodynamics, ruling the time evolution of the locally conserved quantities associated with energy, momentum, particle numbers, and mass \cite{GM84}. In the phases of condensed matter with broken continuous symmetries such as crystals and liquid crystals, there exist further transport processes arising from the Nambu-Goldstone modes, which extend the equations of hydrodynamics \cite{F75,CL95}. A fundamental issue is to deduce these irreversible properties from the underlying microscopic dynamics of particles composing matter.

This issue can be addressed within the local-equilibrium approach, assuming that the time evolution of the system is slow enough for the statistical distribution to remain close to a local Gibbsian distribution expressed in terms of spatiotemporal macrofields such as temperature, chemical potentials, and velocity.  This approach has been developed since the late 1950s and early 1960s, in particular by Mori \cite{M56,M58}, McLennan \cite{McL60,McL61,McL63}, and others \cite{R66,R67,Z74,AP81,OL79,S14,HHNH15,H19,DLW20,MG20,MG21}.  Close to local equilibrium, the transport coefficients can thus be expressed in the form of Green-Kubo formulas \cite{G52,G54,K57}.  In the local-equilibrium approach, the entropy can be introduced and the inferred entropy production plays a key role for identifying the dissipative effects from the microscopic dynamics up to the macroscale.  Furthermore, the ratio between the exact time-evolved probability distribution and the corresponding local-equilibrium distribution obeys a so-called integral fluctuation theorem \cite{S14,HHNH15}.  Within classical microscopic dynamics, the nonnegativity of entropy production can be deduced from the integral fluctuation theorem using Jensen's inequality \cite{S14}.  Since the motion of particles is fundamentally ruled by quantum mechanics, an important problem is to extend these results to the quantum-mechanical formulation of condensed matter physics.  Some results have been obtained in the framework of relativistic quantum field theory using the Heisenberg picture \cite{HHNH15,H19}.  However, the microdynamics of nonrelativistic systems is often formulated using the Schr\"odinger picture, which is closer to classical nonequilibrium statistical mechanics.  In this picture, the local-equilibrium statistical operator should be defined at the current time rather than at the initial time.  

On this basis, we here propose a first-principle derivation of the hydrodynamic equations in the quantum-mechanical framework, where the integral fluctuation theorem is obtained.  Therefrom, the nonnegativity of entropy production can be deduced using the Peierls-Bogoliubov inequality \cite{W78}.  In the Schr\"odinger picture, we establish the equivalence between the local-equilibrium approach and the methods based on projection operators \cite{KG73,G77,G82}.  Using the expansion in the gradients of the macrofields, the transport coefficients are shown to be given by Green-Kubo formulas.  As a consequence of microreversibility, the Onsager-Casimir reciprocal relations are satisfied.  At leading order of the gradient expansion, we can show that the entropy production rate is expressed in terms of the symmetric part of the matrix of transport coefficients and is also nonnegative.  The nonnegativity of the entropy production rate is proved using the quantum version of the Einstein-Helfand formulas \cite{E26,H60}.  The results are general and they apply {\it mutatis mutandis} to fluids and the phases of condensed matter with broken continuous symmetries.

The paper is organized as follows.  In Sec.~\ref{Sec:Definitions}, we present the time evolution of the statistical operator and the microscopic densities of the slow modes.  The local-equilibrium statistical operator is introduced in terms of the microscopic densities and the conjugate fields.  The entropy is defined and functional identities are obtained for the mean values with respect to the local-equilibrium statistical operator.  In Sec.~\ref{Sec:EP-CD}, the time evolution of the exact statistical operator is related to the corresponding local-equilibrium statistical operator in the Schr\"odinger picture.  The quantum integral fluctuation theorem is obtained and the Peierls-Bogoliubov inequality is used to deduce the nonnegativity of entropy production.  The nondissipative and dissipative parts of current densities are identified.  In Sec.~\ref{Sec:GradExp}, their gradient expansion is carried out. General equations are obtained for the time evolution of the conjugate fields and the relationship to projection-operator methods is established.  The transport coefficients are thus given as the coefficients of linear response with respect to the affinities, i.e., the gradients of the conjugate fields.  The transport coefficients are expressed in terms of Green-Kubo formulas.  At leading order in the gradient expansion, the entropy production rate is obtained and shown to be nonnegative, using quantum-mechanical Einstein-Helfand formulas.  In Sec.~\ref{Sec:Appl}, we consider the application of the formalism to multicomponent fluids and condensed matter with broken symmetries.  The conclusion is drawn in Sec.~\ref{Sec:Conclusion}.  In App.~\ref{AppA}, we provide details for the calculations of the formalism.  The details of the derivation are given for multicomponent fluids in App.~\ref{AppB}.  Appendix~\ref{AppC} presents an overview of the derivation for the phases with broken symmetries.

{\it Notations.} Latin letters $a, b, c, \ldots = x, y, z$ correspond to spatial coordinates and Greek letters $\alpha, \beta, \gamma, \ldots$ label the hydrodynamic variables. Unless explicitly stated, Einstein's convention of summation over repeated indices is adopted. $\hbar$ denotes Planck's constant, $k_{\rm B}$ Boltzmann's constant, and $\imath=\sqrt{-1}$.

\section{Microscopic dynamics and statistical operators}
\label{Sec:Definitions}

\subsection{Microscopic densities and time evolution}

The hydrodynamic modes, or slow modes, of the system originate from the fundamental conservation laws (energy, momentum, particle numbers, and mass) and from the Nambu-Goldstone modes associated with the breaking of continuous symmetries. The corresponding microscopic densities are local observables denoted by $\hat{c}^\alpha({\bf r})$, which includes the densities of energy, momentum, particles, and mass, as well as the gradients of the order fields generated by symmetry breaking.  The integrals of these densities $\hat{C}^\alpha \equiv \int\hat{c}^\alpha({\bf r})\, d{\bf r}$ are conserved quantities, such that $[\hat{C}^\alpha,\hat H]=0$, where $\hat H$ is the Hamiltonian operator ruling the time evolution of the many-body wave function.  Here, we assume that the microscopic dynamics is defined in a domain with periodic boundary conditions, which is convenient to describe bulk phases.  In this regard, the system is isolated.

In the Heisenberg picture, the time evolution of the microscopic densities is given by
\begin{align}
\imath \hbar \, \partial_t  \hat{c}^\alpha({\bf r},t) & = [\hat{c}^\alpha({\bf{r}},t),\hat{H}] \, ,
\label{eqs-H-c}
\end{align}
leading to the microscopic local conservation equations
\begin{align}
\partial_t  \, \hat{c}^\alpha({\bf r},t)+\nabla^a\hat{J}^a_{c^\alpha}({\bf r},t)&=0\, ,
\label{eqs-c}
\end{align}
which define the microscopic current densities $\hat{J}^a_{c^\alpha}$.  The microscopic densities and current densities are both defined as Hermitian operators since they are physical observables.  Introducing the Liouvillian superoperator
\begin{align}\label{Liouvillian}
 \widehat{\cal L} \hat A&\equiv \frac{1}{\imath\hbar} \, [\hat H, \hat A]
 \end{align}
acting on any operator $\hat A$, the densities at time $t$ can be expressed as
 \begin{align}\label{micro_c}
 \hat{c}^\alpha({\bf r},t)& \equiv {\rm e}^{-\widehat{\cal L} t} \,  \hat{c}^\alpha({\bf r})= {\rm e}^{\frac{\imath}{\hbar}\hat H t}\,  \hat{c}^\alpha({\bf r})\, {\rm e}^{-\frac{\imath}{\hbar}\hat H t}
 \end{align}
 with similar expressions for the current densities.

\subsection{The exact statistical operator}

We consider a statistical ensemble of copies of the system, which is described by some statistical operator evolving in time according to
 \begin{align}\label{rho(t)}
 \hat{\varrho}_t &= {\rm e}^{\widehat{\cal L} t} \, \hat{\varrho}_0= {\rm e}^{-\frac{\imath}{\hbar}\hat H t}\, \hat{\varrho}_0\, {\rm e}^{\frac{\imath}{\hbar}\hat H t} \, ,
 \end{align}
where $\hat{\varrho}_0$ is the statistical operator at initial time. This time evolution preserves the Hermitian character of the statistical operator, $\hat\varrho_t^{\dagger}=\hat\varrho_t$, as well as its normalization ${\rm tr}\,\hat\varrho_t=1$.  

The mean values of the microscopic densities can thus be equivalently expressed as
\begin{align}
\langle\hat{c}^{\alpha}(\mathbf{r})\rangle_t  = {\rm tr}\, \hat{c}^{\alpha}(\mathbf{r}) \, \hat{\varrho}_t = {\rm tr}\, \hat{c}^{\alpha}(\mathbf{r},t) \, \hat{\varrho}_0
\end{align}
with the notation $\langle\cdot\rangle_t\equiv\text{tr}(\cdot\,\hat{\varrho}_t)$. In the Schr\"odinger picture, the microscopic densities do not depend on time, but the statistical operator does.  In the Heisenberg picture, the microscopic densities evolve in time, but the mean value is taken over the initial statistical operator.  The macroscopic equations are obtained from the ensemble average of Eq.~\eqref{eqs-c} with respect to the initial statistical operator~$\hat{\varrho}_0$, giving
\begin{align}
\partial_t\, \langle\hat{c}^{\alpha}(\mathbf{r})\rangle_t   + \nabla^a \langle\hat{J}^{a}_{c^\alpha}(\mathbf{r})\rangle_t & = 0 \, .\label{eqs-macro-c}
\end{align}
In the rest of the paper, the formalism is developed in the Schr\"odinger picture.

\subsection{The local-equilibrium statistical operator}

The local-equilibrium statistical operator is defined as
\begin{align}
\hat{\varrho}_{\text{leq},\boldsymbol{\lambda}}={\rm e}^{-\hat\varsigma({\boldsymbol{\lambda}})} \qquad \text{with} \qquad \hat\varsigma({\boldsymbol{\lambda}})\equiv{\lambda}^\alpha\ast \hat{c}^\alpha+\Omega(\boldsymbol{\lambda}) \, ,\label{rho-leq}
\end{align}
where $\boldsymbol{\lambda}=({\lambda}^\alpha)$ are inhomogeneous fields conjugated to the densities $\hat{c}^{\alpha}$, the asterisk~$\ast$ is defined by the integration over space, $ f\ast g \equiv \int f({\bf r}) \, g({\bf r})\, d{\bf r}$, and the normalization condition for the local-equilibrium statistical operator gives the Massieu functional $\Omega(\boldsymbol{\lambda}) = \ln {\rm tr}\, \exp\left(- \lambda^\alpha\ast \hat{c}^\alpha \right)$. The mean densities are given by the functional derivatives of the Massieu functional with respect to the conjugate fields:
\begin{align}
c^\alpha(\mathbf{r})&\equiv \langle\hat{c}^\alpha(\mathbf{r})\rangle_{\text{leq},\boldsymbol{\lambda}}=-\frac{\delta \Omega(\boldsymbol{\lambda})}{\delta \lambda^\alpha(\mathbf{r})} \, ,
\label{eq:dCdlambda}
\end{align}
where $\langle\cdot\rangle_{\text{leq},\boldsymbol{\lambda}} \equiv \text{tr}(\cdot \, \hat{\varrho}_{\text{leq},\boldsymbol{\lambda}})$. Therefore, the Massieu functional plays the role of generating functional for the statistical moments of the densities.

 In the local-equilibrium approach \cite{McL61,McL63,R66,Z74}, the entropy of the system is defined in terms of the local-equilibrium statistical operator~\eqref{rho-leq} according to
 \be\label{entropy}
 S \equiv -k_{\rm B} \, {\rm tr}\, \hat\varrho_{{\rm\, leq},\boldsymbol{\lambda}}  \, \ln \hat\varrho_{{\rm\, leq},\boldsymbol{\lambda}} \, .
 \ee
 The entropy is the Legendre transform of the Massieu functional \cite{S14}
\be
S({\bf c}) = k_{\rm B}\, \inf_{\boldsymbol{\lambda}}\left[ {\lambda}^\alpha\ast {c}^\alpha + \Omega(\boldsymbol{\lambda}) \right] ,\label{eq.entropyfunctional}
\ee
so that the entropy is a functional of the mean densities ${\bf c}=(c^\alpha)$ defined by Eq.~\eqref{eq:dCdlambda}.
As a consequence, the conjugate fields are given by the functional derivatives of the entropy functional,
\be\label{eq:SAsecondid}
\lambda^\alpha({\bf r}) =\frac{1}{k_{\rm B}}\, \frac{\delta S({\bf c})}{\delta  {c}^\alpha({\bf r})} \, .
\ee

In order to calculate higher functional derivatives, we need to consider the functional derivative of the local-equilibrium statistical operator itself:
\begin{align}
\frac{\delta  \hat{\varrho}_{\text{leq},\boldsymbol{\lambda}} }{\delta\lambda^\alpha(\mathbf{r})}&=-\int_{0}^{1}dx\, {\rm e}^{-x \hat\varsigma({\boldsymbol{\lambda}})}\, \delta \hat c^\alpha(\mathbf{r}) \, {\rm e}^{x \hat\varsigma({\boldsymbol{\lambda}})} \hat{\varrho}_{\text{leq},\boldsymbol{\lambda}}\, ,
\label{drho/dlambda}
\end{align}
where $\delta \hat c^\alpha \equiv \hat{c}^\alpha-\left\langle\hat{c}^\alpha\right\rangle_{\text{leq},\boldsymbol{\lambda}}$, as shown in Eqs.~\eqref{eq:drholeq} and \eqref{eq:drho} \cite{R66,R67,DLW20}.  In particular, we can define the kernel formed by the density-density correlation functions as
\be
C_{\alpha\beta}({\bf r},{\bf r}') \equiv \langle \delta\hat{c}^\alpha({\bf r}) | \delta\hat{c}^\beta({\bf r}')\rangle_{\boldsymbol{\lambda}} = C_{\beta\alpha}({\bf r}',{\bf r})
\ee
in terms of the Mori product \cite{F75,M62}
\begin{align}
\langle \hat{A}|\hat{B}\rangle_{\boldsymbol{\lambda}}  & \equiv  \text{tr} \int_{0}^{1}dx\, \hat{A}^\dagger \, {\rm e}^{-x \hat\varsigma({\boldsymbol{\lambda}})}\hat{B} \, {\rm e}^{x \hat\varsigma({\boldsymbol{\lambda}})}\hat{\varrho}_{\text{leq},\boldsymbol{\lambda}} = \langle \hat{B}^{\dagger}|\hat{A}^{\dagger}\rangle_{\boldsymbol{\lambda}} \, .
\label{eq:Moriproduct}
\end{align}
This kernel and its inverse are thus given by the second functional derivatives as follows,
\begin{align}
\frac{\delta c^\alpha}{\delta \lambda^\beta} &= - \frac{\delta^2\Omega}{\delta \lambda^\alpha \delta \lambda^\beta} = -\langle \delta\hat{c}^\alpha|\delta\hat{c}^\beta\rangle_{\boldsymbol{\lambda}}\, , 
\label{dc/dlambda}
\end{align}
\begin{align}
 \frac{\delta \lambda^\beta}{\delta c^\alpha}& = \frac{1}{k_{\rm B}}\,  \frac{\delta^2 S}{\delta c^\alpha \delta c^\beta}=-\langle \delta\hat{c}^\alpha|\delta\hat{c}^\beta\rangle^{-1}_{\boldsymbol{\lambda}}\, .
\label{dlambda/dc}
\end{align}
We note that this kernel characterizes the spatial correlation functions of the densities in the local equilibrium defined by the conjugate fields $\boldsymbol{\lambda}$.  Equilibrium statistical mechanics shows that these spatial correlation functions are typically decaying as $\exp(-r/\ell)/r$, where the correlation length $\ell$ is finite away from phase transitions.

The existence of the inverse of the kernel is a condition allowing us to determine in principle the conjugate fields $\boldsymbol{\lambda}=(\lambda^\alpha)$ from the knowledge of the mean densities ${\bf c}=(c^\alpha)$, in particular, during the time evolution.

\subsection{Basic assumptions and consequences}

The formalism of the local-equilibrium approach is based on the key assumption that the system is initially described by some local-equilibrium statistical operator $\hat{\varrho}_0=\hat{\varrho}_{\text{leq},\boldsymbol{\lambda}_0}$, so that Eq.~\eqref{rho(t)} now reads
 \begin{align}\label{rho(t)-leq}
 \hat{\varrho}_t &= {\rm e}^{\widehat{\cal L} t} \, \hat{\varrho}_{\text{leq},\boldsymbol{\lambda}_0}= {\rm e}^{-\frac{\imath}{\hbar}\hat H t}\, \hat{\varrho}_{\text{leq},\boldsymbol{\lambda}_0}\, {\rm e}^{\frac{\imath}{\hbar}\hat H t} \, .
 \end{align}
 Supposing the existence of the inverse kernel~\eqref{dlambda/dc}, the conjugate fields can be defined at any time $t$ on the basis of the requirements that the mean densities calculated with the exact statistical operator are equal to the mean densities with respect to the local-equilibrium statistical operator for the time-evolved conjugate fields $\boldsymbol{\lambda}_t$:
 \be\label{basic_conditions}
 c^\alpha({\bf r},t) \equiv \langle\hat{c}^\alpha(\mathbf{r})\rangle_t = \langle\hat{c}^\alpha(\mathbf{r})\rangle_{\text{leq},\boldsymbol{\lambda}_t} \, .
 \ee
In general, we note that $\hat{\varrho}_t \neq \hat{\varrho}_{\text{leq},\boldsymbol{\lambda}_t}$ for $t\neq 0$, because the equalities~\eqref{basic_conditions} are formulated in a functional space typically much smaller than the space of statistical operators if we consider many-body quantum systems.

 The key assumption~\eqref{rho(t)-leq} has several important consequences. Since the normalization condition ${\rm tr}\, \hat\varrho_t=1$ is always satisfied, the property  $(d/dt){\rm tr}\, \hat\varrho_t|_{t=0}=0$ leads to the following functional identity \cite{OL79,S14},
 \be
{\nabla}^a{\lambda}^\alpha\ast \langle\hat{J}^{a}_{c^\alpha}\rangle_{\text{leq},\boldsymbol{\lambda}} = 0 \, ,
\label{eq:SAfirstid}
\ee
which holds for any conjugate field ${\lambda}^\alpha$, as explained in Eq.~\eqref{eq:SAfirstid-proof}.  Now, the variation of the identity~\eqref{eq:SAfirstid} with respect to $\lambda^\alpha$ gives another functional identity, which reads
\begin{align}
{\nabla}^a\langle\hat{J}^{a}_{c^\alpha}(\mathbf{r})\rangle_{\text{leq},\boldsymbol{\lambda}} &=-\int d\mathbf{r}' \, \langle \delta \hat{c}^{\alpha}(\mathbf{r})| \delta \hat{J}^a_{c^{\beta}}(\mathbf{r}') \rangle_{\boldsymbol{\lambda}}\, {\nabla}'^a {\lambda}^\beta(\mathbf{r}') \label{eq:rel1stid}
\end{align}
 in terms of the Mori product~\eqref{eq:Moriproduct} as shown with Eq.~\eqref{eq:var1idlambda}.  These functional identities play an important role in the following.

\section{Entropy production and current densities}
\label{Sec:EP-CD}

\subsection{Quantum integral fluctuation theorem}

At any time $t$, the entropy~\eqref{entropy} can be expressed as
 \be\label{entropy2}
 S_t = k_{\rm B} \left[\lambda_t^\alpha \ast c_t^\alpha + \Omega(\boldsymbol{\lambda}_t)\right] = -k_{\rm B} \, {\rm tr}\, \hat\varrho_t  \, \ln \hat\varrho_{{\rm\, leq},\boldsymbol{\lambda}_t}
 \ee
by using the definition~\eqref{rho-leq} of the local-equilibrium statistical operator and the requirements~\eqref{basic_conditions} with the notation $c_t^\alpha({\bf r})=c^\alpha({\bf r},t)$ \cite{M56,M58}.  Furthermore, we note that the initial value of the entropy is given by the von~Neumann formula, not only for the initial statistical operator, but also for the exact statistical operator at any time, since the von~Neumann formula gives a time-independent value: $S_0=-k_{\rm B} \, {\rm tr}\, \hat\varrho_0  \, \ln \hat\varrho_0 =-k_{\rm B} \, {\rm tr}\, \hat\varrho_t  \, \ln \hat\varrho_t $.

Using the Liouvillian superoperator~\eqref{Liouvillian}, the time evolution of the exact statistical operator thus reads
\bea
\hat{\varrho}_t &=& {\rm e}^{\widehat{\cal L} t} \,\hat{\varrho}_{\text{leq},\boldsymbol{\lambda}_0} \nonumber\\
&=& \exp\left[-{\lambda}_0^\alpha\ast \hat{c}_{-t}^\alpha-\Omega(\boldsymbol{\lambda}_0)\right] \nonumber \\
&=&\exp\left[-\hat{\varsigma}(\boldsymbol{\lambda}_t)+\hat{\Sigma}_t\right]
\label{rho-Sigma}
\eea
in terms of the operator
\bea
\hat{\Sigma}_t &\equiv&   \int_0^t {\rm d}\tau\,  \partial_\tau \left[{\lambda}_\tau^\alpha\ast\hat{c}^\alpha_{\tau - t} + \Omega(\boldsymbol{\lambda}_\tau)\right] \label{Sigma} \\
&=& \int_0^t {\rm d}\tau \left(\partial_\tau {\lambda}_\tau^\alpha\ast\delta\hat{c}^\alpha_{\tau - t}  + \nabla^a\lambda^\alpha_\tau\ast \delta \hat{J}^a_{c^\alpha,{\tau-t}}\right) , \quad
\label{Sigma2}
\eea
where the last expression is calculated in Eq.~\eqref{Sigma-proof} with
\bea
\delta\hat{c}^\alpha_{\tau-t} &\equiv& \hat{c}_{\tau-t}^\alpha-\langle \hat{c}^{\alpha} \rangle_{\text{leq},\boldsymbol{\lambda}_\tau} \, , \label{delta-c}\\
\delta\hat{J}^a_{c^\alpha,{\tau-t}} &\equiv& \hat{J}^a_{c^\alpha,{\tau-t}} -\langle \hat{J}^a_{c^\alpha}  \rangle_{\text{leq},\boldsymbol{\lambda}_\tau} \, . \label{delta-J}
\eea
We note the crucial role of the operator $\hat{\Sigma}_t$, which is a central quantity in the formalism. Indeed, its mean value with respect to the exact statistical operator $\hat\varrho_t$ is related to the entropy difference between the initial time $t=0$ and the current time $t$ according to
\be\label{Sigma-DS}
\langle \hat\Sigma_t \rangle_t =\text{tr}\left(\hat{\rho}_t\ln\hat{\rho}_t\right) -\text{tr}\left(\hat{\rho}_{\text{leq},\boldsymbol{\lambda}_t}\ln \hat{\rho}_{\text{leq},\boldsymbol{\lambda}_t}\right) = \frac{1}{k_{\rm B}} \left( S_t- S_0\right) ,
\ee
as proved in Eq.~\eqref{mean-Sigma-App}.

Most remarkably, the exact statistical operator can be related to the local-equilibrium statistical operator by the following identity,
\be
\hat{\varrho}_t = \hat{\Xi}_t \, \hat{\varrho}_{\text{leq},\boldsymbol{\lambda}_t} \, ,
\label{rho-Xi-leq}
\ee
where
\bea
\hat{\Xi}_t &\equiv& \hat{\varrho}_t \, \hat{\varrho}_{\text{leq},\boldsymbol{\lambda}_t}^{-1}={\rm e}^{-\hat{\varsigma}(\boldsymbol{\lambda}_t)+\hat{\Sigma}_t}\, {\rm e}^{\hat{\varsigma}(\boldsymbol{\lambda}_t)} \nonumber\\
&=& 1+\int_0^1dx \, {\rm e}^{x\left[-\hat\varsigma({\boldsymbol{\lambda}}_t)+\hat{\Sigma}_t\right]} \, \hat{\Sigma}_t \, {\rm e}^{x\hat\varsigma({\boldsymbol{\lambda}}_t)} \, , \label{eq:Xi}
\eea
as Eq.~\eqref{rho-Xi-leq-App} shows.
Therefore, the ensemble average of any operator $\hat{A}$ can be expressed at any time in terms of an ensemble average with respect to the corresponding local-equilibrium statistical operator, providing the inclusion of the operator $ \hat{\Xi}_t$ as follows,
\begin{align}
\langle\hat{A}\rangle_t & = \langle\hat{A}\, \hat{\Xi}_t\rangle_{\text{leq},\boldsymbol{\lambda}_t}\, ,\label{eq:SAthirdid}
\end{align}
which is a further fundamental identity of the formalism. Next, the choice $\hat{A}=\hat{\Xi}_t^{-1}$ in Eq.~\eqref{eq:SAthirdid} gives a quantum version of the integral fluctuation theorem
\begin{align}\label{QIFT}
\left\langle {\rm e}^{-\hat\varsigma({\boldsymbol{\lambda}}_t)}\, {\rm e}^{\hat{\varsigma}(\boldsymbol{\lambda}_t)-\hat{\Sigma}_t}\right\rangle_t=1\, .
\end{align}
Now, we have the Peierls-Bogoliubov inequality \cite{W78}
\be\label{QJI}
\left\langle {\rm e}^{-\hat\varsigma({\boldsymbol{\lambda}}_t)}\,  {\rm e}^{\hat\varsigma({\boldsymbol{\lambda}}_t)-\hat\Sigma_t}\right\rangle_t \ge {\rm e}^{-\langle\hat\Sigma_t\rangle_t} \, ,
\ee
which is obtained with Eqs.~\eqref{PB-ineq}-\eqref{pre-QJI}.
By combining the quantum integral fluctuation theorem~\eqref{QIFT} with the Peierls-Bogoliubov inequality~\eqref{QJI}, we deduce that $\langle \hat\Sigma_t \rangle_t  \ge 0$.  As a consequence of Eq.~\eqref{Sigma-DS}, we thus obtain the inequality
\be\label{Inequality-DS}
S_t- S_0 = k_{\rm B} \langle \hat\Sigma_t \rangle_t \ge 0 \, .
\ee
Since the system is here assumed to be isolated, this inequality can be interpreted as the nonnegativity of the entropy production during the time interval $[0,t]$ with $t\ge 0$.

\subsection{Reversible and dissipative current densities}
\label{Subsec:rev-dis-cd}

The time derivative of the entropy~\eqref{entropy} is given by
\begin{align}
\frac{1}{k_{\rm B}}\, \frac{dS}{dt} = \nabla^a\lambda^{\alpha}_t\ast  \langle \hat{J}^a_{c^{\alpha}}\rangle_{t} \, ,\label{eq:epr}
 \end{align}
 as shown in Eq.~\eqref{dS/dt-proof}. The mean values of the current densities can always be decomposed as
\be
\langle \hat{J}^a_{c^\alpha}(\mathbf{r})\rangle_t =\bar{J}^a_{c^\alpha} (\mathbf{r},t) + \mathcal{J}^a_{c^\alpha}(\mathbf{r},t) \, ,
\label{mean_crnt_dens}
\ee
where
\bea
\bar{J}^a_{c^\alpha} (\mathbf{r},t) &\equiv& \langle\hat{J}^{a}_{c^\alpha}(\mathbf{r})\rangle_{\text{leq},\boldsymbol{\lambda}_t} \qquad\qquad\mbox{and} \label{eq:rev-cd} \\
\mathcal{J}^a_{c^\alpha}(\mathbf{r},t) &\equiv& \langle\delta\hat{J}^{a}_{c^\alpha}(\mathbf{r})\rangle_t=\langle\delta\hat{J}^{a}_{c^\alpha}(\mathbf{r})\,(\hat{\Xi}_t-1)\rangle_{\text{leq},\boldsymbol{\lambda}_t} \qquad \label{eq:dis-cd}
\eea 
with $\delta\hat{J}^{a}_{c^\alpha}(\mathbf{r}) \equiv \hat{J}^{a}_{c^\alpha}(\mathbf{r}) - \langle\hat{J}^{a}_{c^\alpha}(\mathbf{r})\rangle_{\text{leq},\boldsymbol{\lambda}_t}$, since $\langle\delta\hat{J}^{a}_{c^\alpha}(\mathbf{r})\rangle_t=\langle\delta\hat{J}^{a}_{c^\alpha}(\mathbf{r})\,\hat{\Xi}_t\rangle_{\text{leq},\boldsymbol{\lambda}_t}$ by Eq.~\eqref{eq:SAthirdid} and $\langle\delta\hat{J}^{a}_{c^\alpha}(\mathbf{r})\rangle_{\text{leq},\boldsymbol{\lambda}_t} =0$.

Because of the identity~\eqref{eq:SAfirstid}, we have that ${\nabla}^a{\lambda}^\alpha\ast \bar{J}^a_{c^\alpha} = 0$, so that the time derivative of the entropy reduces to
\be
\frac{1}{k_{\rm B}}\, \frac{dS}{dt} = {\nabla}^a{\lambda}^\alpha_t\ast   \mathcal{J}^{a}_{c^\alpha}(t) \, . 
\label{eq:entropyrate-dissip_crnt}
\ee
Combining with the inequality~\eqref{Inequality-DS}, we have that
\begin{align}
\Delta S_t = S_t-S_0 = k_{\rm B} \int_0^t {\nabla}^a{\lambda}^\alpha_\tau \ast   \mathcal{J}^{a}_{c^\alpha}(\tau) \, d\tau \ge 0\, ,
\label{eq:entropyprod}
\end{align}
which is the entropy production during the time interval $[0,t]$, if the system is isolated.  Accordingly, the part~\eqref{eq:rev-cd} of the mean current density~\eqref{mean_crnt_dens} preserves the entropy and can be interpreted as the reversible or dissipativeless part of the corresponding current density.  The dissipative part of the mean current density is thus given by Eq.~\eqref{eq:dis-cd}.  Therefore, the identity~\eqref{eq:SAfirstid} of the formalism leads to the identification of the reversible and dissipative parts of the current densities.

According to Eq.~\eqref{eq:Xi}, the dissipative part of the current densities defined in Eq.~\eqref{eq:dis-cd} becomes a function of $ \hat{\Sigma}_t$ and is given by an ensemble average with respect to the local-equilibrium statistical operator:
\be
 {\mathcal{J}}^a_{c^{\alpha}}(\mathbf{r},t) = \left\langle\delta\hat{J}^{a}_{c^\alpha}(\mathbf{r})\int_0^1dx\, {\rm e}^{x\left[-\hat\varsigma({\boldsymbol{\lambda}}_t)+\hat{\Sigma}_t\right]} \hat{\Sigma}_t \, {\rm e}^{x\hat\varsigma({\boldsymbol{\lambda}}_t)}\right\rangle_{\text{leq},\boldsymbol{\lambda}_t} . 
 \label{eq:dis-cd_formula}
\ee
The macroscopic local conservation equations can thus be written as
\begin{align}\label{bal-eq-c}
\partial_t\, c^{\alpha}  + \nabla^a\left(\bar{J}^{a}_{c^\alpha}+{\mathcal J}^{a}_{c^\alpha}\right)  = 0 
\end{align}
in terms of the reversible and dissipative parts of the current densities given by Eqs.~\eqref{eq:rev-cd} and \eqref{eq:dis-cd_formula}, which is a general and exact result of the formalism.

In nonrelativistic systems, Galilean invariance can be used to express the reversible part~\eqref{eq:rev-cd} of the mean current densities in terms of the velocity field, which is defined as the velocity of the mass center for each element of the continuous medium of interest:
\be
v^a({\bf r},t) \equiv \frac{\langle \hat{g}^a({\bf r})\rangle_t}{\langle \hat{\rho}({\bf r})\rangle_t} \, ,
\label{velocity-dfn}
\ee
where $\hat{\rho}$ and $\hat{g}^a$ denote the microscopic densities of mass and linear momentum, respectively.  These quantities obey the local conversation equation
\be
\partial_t \, \hat{\rho} + \nabla^a \hat{g}^a = 0 \, ,
\label{mass-bal-eq}
\ee
so that the corresponding macroscopic equation is the continuity equation
\be
\partial_t \, \rho + \nabla^a(\rho\, v^a) = 0 \, ,
\label{cont-eq}
\ee
where $\rho({\bf r},t)\equiv\langle \hat{\rho}({\bf r})\rangle_t$.  As a direct consequence, the mean current density of mass is entirely given by its reversible part $\bar{J}_{\rho}^a=\rho\, v^a$, while its dissipative part is equal to zero ${\mathcal{J}}_{\rho}^a=0$.  Since the mass density is the sum over the particle species $1\leq k \leq \nu$ of the particle densities $\hat{n}_k$ multiplied by their mass $m_k$, $\hat{\rho}=\sum_k m_k \hat{n}_k$, the latter relation implies that $\sum_k m_k {\mathcal{J}}_{n_k}^a=0$ for the dissipative parts of the particle current densities.  Therefore, the local conservation of mass is a constraint eliminating one dissipative current density in the system.

The reversible parts of the other current densities can also be expressed in terms of the velocity field~\eqref{velocity-dfn}, as explained for multicomponent fluids in App.~\ref{AppB}.

We note that the nonnegativity of the entropy production~\eqref{eq:entropyprod} does not imply the inequality $dS/dt\ge 0$  \cite{R66}, so that further assumptions are required before $dS/dt$ might be interpreted as the entropy production rate, as discussed in Sec.~\ref{Sec:GradExp}.

\section{Gradient expansion}
\label{Sec:GradExp}

The calculation of the mean current densities~\eqref{mean_crnt_dens} is achieved by using an expansion in the gradients of the macrofields.  For continuous media under typical macroscopic nonequilibrium conditions, these gradients have a spatial scale much larger than the characteristic length of the spatial correlations among the particles.  Under such conditions, the macroscopic movements can be described with the leading terms of the gradient expansion in the continuous medium of interest.  Usually, the reversible part~\eqref{eq:rev-cd} of the mean current densities does not depend on the gradients, while their dissipative part~\eqref{eq:dis-cd} is linear in the gradients.  Accordingly, the dissipative current densities are proportional to the gradients with linear response coefficients defining the transport coefficients.  At leading order of the gradient expansion, the time derivative of the entropy~\eqref{eq:entropyrate-dissip_crnt} is thus given by a quadratic form of the gradients.
In this section, we present the derivation of these results, applicable to systems with modes originating from the fundamental conservation laws and the breaking of continuous symmetries. 

\subsection{Local thermodynamics}

The first step is the identification of the conjugate fields $\lambda^\alpha$. For this purpose, the entropy functional \eqref{eq.entropyfunctional} should be expressed as
\begin{align}\label{S-local-s}
 S({\bf c}) = \int s({\bf c}) \, d{\bf r} + O(\nabla^2) \, ,
\end{align}
in terms of the entropy density $s({\bf c})$ obeying the local Gibbs relation, $ds =k_{\rm B}\, \lambda^\alpha d{ c^{\alpha}}$.
This thermodynamic quantity can be calculated using equilibrium statistical mechanics and assuming that local equilibrium holds in every volume element of the continuous medium.  As a consequence, the conjugate fields are given by Eq.~\eqref{eq:SAsecondid}, so that $\lambda^\alpha = k_{\rm B}^{-1}(\partial s/\partial c^\alpha) + O(\nabla^2)$ up to possible corrections going as the square of the gradients.

According to the identity~\eqref{eq:SAfirstid}, the reversible part of the current densities~\eqref{eq:rev-cd} does not contribute to the time derivative of the entropy~\eqref{eq:entropyrate-dissip_crnt}, so that entropy is preserved by the macroscopic hydrodynamic equations truncated to the reversible parts of the current densities.  These are the Eulerian hydrodynamic equations, which thus describe adiabatic (i.e., dissipativeless) processes in the continuous medium.
As explained in Subsec.~\ref{Subsec:rev-dis-cd}, the reversible part of the current densities can be expressed in terms of the velocity field defined by Eq.~\eqref{velocity-dfn}, as well as thermodynamic quantities such as the hydrostatic pressure and the heat capacity, which are defined using local thermodynamics and the entropy density introduced in Eq.~\eqref{S-local-s}.

\subsection{Dissipative current densities at leading order}

We proceed with the calculation of the dissipative current densities~\eqref{eq:dis-cd_formula} at leading order in the gradient expansion.  From Eq.~\eqref{Sigma2}, we have $\hat{\Sigma}_t\sim O(\nabla)$, so that the dissipative current densities become
\begin{align}
& {\mathcal{J}}^a_{c^{\alpha}}(\mathbf{r},t) \notag\\
&=  \left\langle\delta\hat{J}^{a}_{c^\alpha}(\mathbf{r})\int_0^1dx\, {\rm e}^{-x\hat\varsigma({\boldsymbol{\lambda}}_t)} \, \hat{\Sigma}_t \, {\rm e}^{x\hat\varsigma({\boldsymbol{\lambda}}_t)}\right\rangle_{\text{leq},\boldsymbol{\lambda}_t} + O(\hat{\Sigma}_t^2) \notag\\
&=  \left\langle\delta\hat{J}^{a}_{c^\alpha}(\mathbf{r}) \Big\vert \hat{\Sigma}_t \right\rangle_{\boldsymbol{\lambda}_t} + O(\hat{\Sigma}_t^2) 
\label{dfn-J-cal}
\end{align}
with the Mori product~\eqref{eq:Moriproduct}.

In order to obtain the dissipative current densities, we must evaluate  $\hat{\Sigma}_t$ at leading order in the gradients. With this aim, we consider its expression~\eqref{Sigma2}.  The first terms in the integrand $\partial_\tau {\lambda}_\tau^\alpha\ast\delta\hat{c}^\alpha$ can be obtained starting from the dissipativeless Eulerian hydrodynamic equations for the mean densities $c^\alpha$, since the conjugate fields $\lambda^\alpha$ are functions of these densities within the framework of local thermodynamics \cite{S14,MG20}.  

Remarkably, the dissipativeless equations for the conjugate fields $\lambda^\alpha$ can also be obtained in terms of a projection operator by using the functional identities~\eqref{eq:SAfirstid} and \eqref{eq:rel1stid} of the formalism.  Indeed, the time derivative of the conjugate fields are first given by $\partial_t\lambda^\alpha=(\delta\lambda^\alpha/\delta c^\beta)\ast \partial_tc^\beta$.  Using the second functional derivatives~\eqref{dlambda/dc} and the macroscopic equations~\eqref{eqs-macro-c}, we thus have
\begin{align}
\partial_t\lambda_t^\alpha(\mathbf{r})&=\int d\mathbf{r}' \, \langle \delta \hat{c}^{\alpha}(\mathbf{r})|\delta\hat{c}^\beta(\mathbf{r}') \rangle^{-1}_{\boldsymbol{\lambda}_t} \ \nabla'^a \langle \hat J^a_{c^\beta}({\bf r}')\rangle_t \, .\label{eq:dlambdadt}
\end{align}
Decomposing the mean current densities according to Eq.~\eqref{mean_crnt_dens} and using the functional identity~\eqref{eq:rel1stid} for ${\nabla'}^a \bar{J}^a_{c^\beta}$ give
\begin{widetext}
\begin{align}
\partial_\tau {\lambda}_\tau^\alpha \ast\delta\hat{c}^\alpha = &-\iiint d\mathbf{r}\, d\mathbf{r}' \, d\mathbf{r}'' \, \delta\hat{c}^\alpha(\mathbf{r}) \, \langle \delta \hat{c}^{\alpha}(\mathbf{r})|\delta\hat{c}^\beta(\mathbf{r}') \rangle^{-1}_{\boldsymbol{\lambda}_\tau} \, \langle \delta \hat{c}^{\beta}(\mathbf{r}')| \delta \hat{J}^a_{c^{\gamma}}(\mathbf{r}'') \rangle_{\boldsymbol{\lambda}_\tau} \, {\nabla}''^a {\lambda}^\gamma_\tau(\mathbf{r}'') \notag\\
	&+\iint d\mathbf{r}\, d\mathbf{r}' \, \delta\hat{c}^\alpha(\mathbf{r})\, \langle \delta \hat{c}^{\alpha}(\mathbf{r})|\delta\hat{c}^\beta(\mathbf{r}') \rangle^{-1}_{\boldsymbol{\lambda}_\tau} \, {\nabla}'^a {\mathcal{J}}^a_{c^{\beta}}(\mathbf{r}',\tau)\, .\label{eq:decf}
\end{align}
\end{widetext}
The first term can be expressed in terms of the  projection operator defined as
\bea
&& \widehat{\mathcal{P}}_{\boldsymbol{\lambda}}\hat{X} \equiv\delta\hat{c}^\alpha \ast\frac{\delta }{\delta c^\alpha}\langle \hat{X}\rangle_{\text{leq},\boldsymbol{\lambda}} \label{eq:pop} \\
&&= \iint d\mathbf{r}\, d\mathbf{r}' \, \delta\hat{c}^\alpha(\mathbf{r}) \, \langle \delta \hat{c}^{\alpha}(\mathbf{r})|\delta\hat{c}^\beta(\mathbf{r}') \rangle^{-1}_{\boldsymbol{\lambda}} \, \langle \delta \hat{c}^{\beta}(\mathbf{r}')| \hat{X}\rangle_{\boldsymbol{\lambda}}
\nonumber
\eea
for an arbitrary operator $\hat X$, as shown in Eq.~\eqref{eq:dlambda-projection}.  Since ${\mathcal{J}}^a_{c^{\beta}}\sim O(\hat{\Sigma}_t)$ and $\hat{\Sigma}_t\sim O(\nabla)$, the second term of Eq.~\eqref{eq:decf} is of higher order in the gradient expansion and can be neglected at leading order of the expansion, so that
\begin{align}
\partial_\tau {\lambda}_\tau^\alpha \ast\delta\hat{c}^\alpha &=- \nabla^a{\lambda_\tau^\alpha}\ast \widehat{\mathcal{P}}_{\boldsymbol{\lambda}_\tau}\delta \hat{J}^a_{c^{\alpha}}+ O(\nabla^2)\, .\label{eq:decfprop}
\end{align}
The equivalence of this approach with the methods based on projection operators~\cite{KG73,G77,G82} is therefore established. This equivalence also holds for classical systems.  We stress that the matrix elements $\langle \delta \hat{c}^{\alpha}(\mathbf{r})|\delta\hat{c}^\beta(\mathbf{r}') \rangle_{\boldsymbol{\lambda}_\tau}$ and $\langle \delta \hat{c}^{\alpha}(\mathbf{r})|\delta\hat{J}^a_{c^\beta}(\mathbf{r}') \rangle_{\boldsymbol{\lambda}_\tau}$ are the central building blocks of the approach based on projection operators. These matrix elements can be directly obtained from the reversible current densities and the dissipativeless equations for the conjugate fields $\lambda^\alpha$ by using Eqs.~\eqref{eq:rel1stid} and~\eqref{eq:dlambdadt}. 

Substituting the result~\eqref{eq:decfprop} into the integrand of $\hat{\Sigma}_t$, the latter is expressed as
\begin{align}
\hat{\Sigma}_t &=\int_0^td\tau\, \mathcal{A}^{a}_{c^\alpha}(\tau)\ast\delta\hat{J}^{\prime a}_{c^\alpha}(\tau-t)+ O(\nabla^2) \, ,
\end{align}
where the gradients of the conjugate fields define thermodynamic forces or affinities corresponding to the densities~$c^\alpha$
\be\label{affinity-dfn}
\mathcal{A}^{a}_{c^\alpha} \equiv \nabla^a\lambda^\alpha \, ,
\ee
and the prime current densities are defined as
\be\label{prime-cd}
\delta\hat{J}^{\prime a}_{c^\alpha}\equiv\delta\hat{J}^{a}_{c^\alpha}-\widehat{\mathcal{P}}_{\boldsymbol{\lambda}_\tau}\delta\hat{J}^{a}_{c^\alpha} \, .
\ee
We note that the projection operator~\eqref{eq:pop} satisfies the property $\langle \widehat{\mathcal{P}}_{\boldsymbol{\lambda}}\hat{X}\rangle_t=0$, because $\langle\delta\hat{c}^\alpha\rangle_t=0$. 
Consequently, the dissipative part~\eqref{eq:dis-cd} of the current densities can be written as $\mathcal{J}^a_{c^\alpha}(\mathbf{r},t)\equiv \langle\delta\hat{J}^{a}_{c^\alpha}(\mathbf{r})\rangle_t= \langle\delta\hat{J}^{\prime a}_{c^\alpha}(\mathbf{r})\rangle_t$, so that $\delta\hat{J}^{a}_{c^\alpha}$ can be replaced by $\delta\hat{J}^{\prime a}_{c^\alpha}$ in Eqs.~\eqref{eq:dis-cd}, \eqref{eq:dis-cd_formula}, and \eqref{dfn-J-cal} \cite{HHNH15,H19}.

The dissipative current densities are finally obtained as
\begin{widetext}
\begin{align}
{\mathcal{J}}^a_{c^{\alpha}}(\mathbf{r},t) & = \int_0^t d\tau \int d{\bf r'}\, \left\langle \delta\hat{J}^{\prime a}_{c^\alpha}(\mathbf{r})\Big\vert \delta\hat{J}^{\prime b}_{c^\beta}(\mathbf{r}',\tau-t)\right\rangle_{\boldsymbol{\lambda}_t}  \mathcal{A}^{b}_{c^\beta}(\mathbf{r}',\tau)+O(\nabla^2)
\label{dis-cd-aff}
\end{align}
\end{widetext}
with the Mori product~\eqref{eq:Moriproduct}.

\subsection{Green-Kubo formulas}

The current correlation functions appearing in the integrand of Eq.~\eqref{dis-cd-aff} are decaying to zero on the spatiotemporal scales that characterize the structure and dynamics of the phase of interest.  The macrofields are assumed to vary in space and time over scales that are larger than the correlation time and the correlation length of the current correlation functions.  Accordingly, the affinities $\mathcal{A}^{b}_{c^\beta}(\mathbf{r}',\tau)$ can be replaced by $\mathcal{A}^{b}_{c^\beta}(\mathbf{r},t)$ in Eq.~\eqref{dis-cd-aff}.

As a further consequence, the conjugate fields ${\lambda}^\alpha_{\tau}$ may be supposed to be quasi uniform on the scales of the current correlation functions. Under such circumstances, the local-equilibrium statistical operator $\hat{\varrho}_{{\rm leq},\boldsymbol{\lambda}_\tau}$ can be replaced by the equilibrium grand canonical statistical operator
\be
\hat{\varrho}_{\rm eq} = {\rm e}^{-\hat\varsigma(\boldsymbol{\lambda}_{\rm eq})} ={\rm e}^{-(k_{\rm B}T)^{-1}(\hat H - \sum\nolimits_{k=1}^{c} \mu_{k0}\hat N_k)-\Omega_{\rm eq}} \, , 
\ee
where $N_k$ is the number of particles of species $k$ in the system and $\mu_{k0}$ is the corresponding chemical potential in the frame at rest with the system.  This equilibrium statistical operator commutes with the Hamiltonian operator $[\hat H,\hat{\varrho}_{\rm eq}]=0$, because the particle number operators do so, $[\hat H,\hat N_k]=0$. Therefore, the equilibrium statistical operator is time invariant, so that the equilibrium statistical properties are stationary and the current correlation functions can be transformed as
\bea
\left\langle \delta\hat{J}^{\prime a}_{c^\alpha}(\mathbf{r},0)\Big\vert \delta\hat{J}^{\prime b}_{c^\beta}(\mathbf{r}',\tau-t)\right\rangle_{\boldsymbol{\lambda}_{\rm eq}} \nonumber\\
= \left\langle \delta\hat{J}^{\prime a}_{c^\alpha}(\mathbf{r},t-\tau)\Big\vert \delta\hat{J}^{\prime b}_{c^\beta}(\mathbf{r}',0)\right\rangle_{\boldsymbol{\lambda}_{\rm eq}} \, .
\eea

Over spatial scales much larger than the distance between the particles, the material properties can be defined by averaging over space. The microscopic global currents are introduced as 
\be
\delta\hat{ \mathbb J}_{c^\alpha}^{a}(t) \equiv \int_V \delta\hat{J}_{c^\alpha}^{a}({\bf r},t)\, d{\bf r}
\label{tot_micro_crnt}
\ee
for the unprime and prime quantities.

Putting everything together, the dissipative current densities become
\be
 {\mathcal{J}}^a_{c^{\alpha}}(\mathbf{r},t)  \simeq \frac{1}{V}
\int_0^{\infty} d\tau \,  \left\langle\delta\hat{ \mathbb J}_{c^\alpha}^{\prime a}(\tau) \Big\vert \delta\hat{\mathbb J}^{\prime b}_{c^\beta}(0) \right\rangle_{\boldsymbol{\lambda}_{\rm eq}} \mathcal{A}^{b}_{c^\beta}(\mathbf{r},t) \,.
\label{eq:DissCurrents-7}
\ee

In the canonical equilibrium ensemble, where the particle numbers $\{ N_k\}$ take given and fixed values,  the dissipative current densities can thus be expressed at leading order as
\be\label{J=LA}
 {\mathcal{J}}^a_{c^{\alpha}}(\mathbf{r},t) ={\cal L}^{ab}_{\alpha\beta}\, \mathcal{A}^{b}_{c^\beta}(\mathbf{r},t)
 \ee
 in terms of the linear response coefficients given by the Green-Kubo formulas
 \be
 {\cal L}^{ab}_{\alpha\beta}\equiv \frac{k_{\rm B}T}{V} \int_0^{\infty} d\tau \int_0^{(k_{\rm B}T)^{-1}} d\vartheta \, \left\langle  \delta\hat{ \mathbb J}_{c^\alpha}^{\prime a}(\tau) \, \delta\hat{{\mathbb J}}^{\prime b}_{c^\beta}(\imath\hbar\vartheta)\right\rangle_{\rm eq} \, ,
\label{eq:GK_formula}
\ee
where $\delta\hat{ \mathbb J}_{c^\alpha}^{\prime a}(t) = {\rm e}^{\frac{\imath}{\hbar}\hat H t} \,\delta\hat{ \mathbb J}_{c^\alpha}^{\prime a}(0) \, {\rm e}^{-\frac{\imath}{\hbar}\hat H t}$  for $t=\tau$ and $t=\imath\hbar\vartheta$ with $\vartheta\equiv (k_{\rm B}T)^{-1} x$.  The limit of infinite volume~$V$ should be taken to obtain the transport coefficients as properties holding at the macroscale.  

Since the Hamiltonian ruling the microscopic dynamics has the time-reversal symmetry, the Onsager-Casimir reciprocal relations are satisfied:
\be\label{OCRR}
{\cal L}^{ab}_{\alpha\beta} =\epsilon_{\alpha}\, \epsilon_{\beta}\, {\cal L}^{ba}_{\beta\alpha} \, ,
\ee
where $\epsilon_{\alpha}=\pm 1$ is the parity of $\delta\hat{ \mathbb J}_{c^\alpha}^{\prime a}$ under time reversal  (and there is no Einstein's summation for these parities).  The relations~\eqref{OCRR} are proved by Eqs.~\eqref{phi(t)}-\eqref{L-int-phi}.

In the classical limit where the thermal de Broglie wave length of the particles is significantly smaller than the mean distance between neighboring particles, the Green-Kubo formulas~\eqref{eq:GK_formula} take their classical form,
 \be
\lim_{\hbar\to 0} {\cal L}^{ab}_{\alpha\beta} = \frac{1}{V} \int_0^{\infty} d\tau \, \left\langle  \delta\hat{ \mathbb J}_{c^\alpha}^{\prime a}(\tau) \, \delta\hat{{\mathbb J}}^{\prime b}_{c^\beta}(0)\right\rangle_{\rm eq} \, ,
\label{eq:class-GK-formula}
\ee
as expected.

\subsection{Entropy production rate}

At leading order of the gradient expansion, the dissipative current densities are linear functions of the affinities, so that the time derivative of the entropy is a quadratic form of the affinities.  Consequently, the antisymmetric part of the matrix of linear response coefficients~\eqref{eq:GK_formula} does not contribute to the time derivative of the entropy, which is solely given in terms of its symmetric part
\be
{\cal L}^{ab\, {\rm S}}_{\alpha\beta} \equiv \frac{1}{2} \left(  {\cal L}^{ab}_{\alpha\beta}+ {\cal L}^{ba}_{\beta\alpha}\right).
 \label{symm-L}
\ee
Using Eqs.~\eqref{J=LA}, and~\eqref{symm-L}, the time derivative~\eqref{eq:entropyrate-dissip_crnt} for the entropy can thus be written as
\be
\frac{1}{k_{\rm B}}\, \frac{dS}{dt} = {\cal L}^{ab\, {\rm S}}_{\alpha\beta}\, \mathcal{A}^{a}_{c^\alpha}(t)\ast  \mathcal{A}^{b}_{c^\beta}(t) \, .
\label{eq:epr_tr}
\ee

Because of the Onsager-Casimir reciprocal relations~\eqref{OCRR}, the symmetric part~\eqref{symm-L} can be expressed as
\be
{\cal L}^{ab\, {\rm S}}_{\alpha\beta} \equiv \frac{1}{2} \left(  1+ \epsilon_{\alpha}\, \epsilon_{\beta}\right) {\cal L}^{ab}_{\alpha\beta} \, ,
 \label{symm-L-OCRR}
\ee
showing that the transport coefficients coupling processes with opposite parities under time reversal do not contribute to the time derivative of the entropy~\eqref{eq:epr_tr}.

Now, if we introduce the Helfand moments \cite{H60} as
\be\label{Helfand-H}
\hat{\mathbb H}_{c^\alpha}^{\prime a}(t,\vartheta) \equiv \int_{\imath\hbar\vartheta/2}^{t+\imath\hbar\vartheta/2} \delta\hat{\mathbb J}_{c^\alpha}^{\prime a}(\tau) \, d\tau \, ,
\ee
the symmetrized transport coefficients~\eqref{symm-L} can be equivalently obtained using the Einstein-Helfand formulas
\be
{\cal L}^{ab\, {\rm S}}_{\alpha\beta} = k_{\rm B}T \int_0^{(k_{\rm B}T)^{-1}}  \Upsilon^{ab}_{\alpha\beta}(\vartheta) \, d\vartheta
\label{EH-formula}
\ee
with
\be   
\Upsilon^{ab}_{\alpha\beta}(\vartheta) \equiv \lim_{t\to\infty} \frac{1}{2tV} \, \langle \hat{\mathbb H}_{c^\alpha}^{\prime a\dagger}(t,\vartheta)\, \hat{\mathbb H}_{c^\beta}^{\prime b}(t,\vartheta)\rangle_{\rm eq} \, , 
\label{Phi-dfn}
\ee
as shown with Eq.~\eqref{XXX}. Since Eq.~\eqref{Phi-dfn} defines a nonnegative quadratic form, the symmetric matrix~\eqref{symm-L} has the same property, $({\cal L}^{ab\, {\rm S}}_{\alpha\beta})\ge 0$, which proves the nonnegativity of Eq.~\eqref{eq:epr_tr}: $dS/dt\ge 0$.  Therefore, the time derivative of the entropy can be interpreted as the entropy production rate at leading order of the gradient expansion in agreement with the second law of thermodynamics.

We note that phenomenological considerations might result into different choices for the affinities \cite{GM84}. Nevertheless, the phenomenological affinities ${\mathcal A}^a_{c^\alpha,{\rm ph}}$ are usually related by linear transformations to the affinities~\eqref{affinity-dfn} that we have here defined as the gradients of the conjugate fields:
\be\label{ph_Aff}
{\mathcal A}^a_{c^\alpha} = {\mathcal M}^{ab}_{\alpha\beta}\,  {\mathcal A}^b_{c^\beta,{\rm ph}} \, .
\ee
The corresponding phenomenological dissipative current densities are thus given by
\be
 {\mathcal J}^b_{c^\beta,{\rm ph}}=  {\mathcal J}^a_{c^\alpha} \, {\mathcal M}^{ab}_{\alpha\beta}\, ,
 \ee
 so that the entropy production rate has the same quadratic form as in Eq.~\eqref{eq:epr_tr}, but with the phenomenological coefficients
\be
{\mathcal L}^{ab}_{\alpha\beta,{\rm ph}}= {\mathcal L}^{a'b'}_{\alpha'\beta'}\, {\mathcal M}^{a'a}_{\alpha'\alpha}\,  {\mathcal M}^{b'b}_{\beta'\beta} \, .
\ee

\section{Applications}
\label{Sec:Appl}

The previous results apply after identifying the microscopic densities of the slow modes in the system of interest.

\subsection{Multicomponent fluids}

In nonreactive multicomponent fluids, the slow modes are associated with the fundamental conservation laws of energy, linear momentum, and particle numbers for the different particle species composing the system.  Accordingly, the microscopic densities are given by
\be
\hat{c}^\alpha = (\hat{e},\hat{g}^a,\hat{n}_k) \, ,
\label{fluid-c}
\ee
where $\hat{e}$ is the energy density, $\hat{g}^a$ with $a=x,y,z$ are the three Cartesian components of linear momentum, and $\hat{n}_k$ with $1\leq k \leq \nu$ are the particle densities for the $\nu$ particle species in the system. These densities obey the local conservation equations~\eqref{eqs-c} with the corresponding microscopic current densities, which can be deduced from the Hamiltonian operator of the system according to Eq.~\eqref{eqs-H-c}.  The details of the calculations are presented in App.~\ref{AppB}.  The local-equilibrium statistical operator can thus be defined by Eq.~\eqref{rho-leq} with the microscopic densities~\eqref{fluid-c} and the conjugate fields $\lambda^\alpha=(\lambda_e,\lambda_{g^a},\lambda_{n_k})$.

In multicomponent fluids, the transport properties are the shear and bulk viscosities, the heat conductivity, and the diffusion and thermodiffusion coefficients.  All these coefficients are given by Green-Kubo formulas.  The thermodiffusion and cross-diffusion coefficients obey Onsager-Casimir reciprocal relations~\eqref{OCRR} with $\epsilon_\alpha\epsilon_\beta=1$, so that the matrix of transport coefficients is symmetric ${\mathcal L}^{ab}_{\alpha\beta}={\mathcal L}^{ab\, {\rm S}}_{\alpha\beta}$.  Therefore, the Green-Kubo formulas~\eqref{eq:GK_formula} are equivalent to Einstein-Helfand formulas~\eqref{EH-formula} and all the aforementioned transport properties contribute to the entropy production rate~\eqref{eq:epr_tr}.

\subsection{Phases with broken continuous symmetries}

In condensed matter with broken continuous symmetries, the slow modes include not only the modes associated with the fundamentally conserved quantities, but also the Nambu-Goldstone modes emerging from symmetry breaking.  The number of these extra modes is equal to the number of continuous symmetries that are broken.  In liquid crystals, rotational symmetries can be broken \cite{F75,CL95}.  In crystals, where the symmetries under three-dimensional spatial translations are broken, there are three Nambu-Goldstone modes emerging from the crystalline long-range order.  Such an order is described by some microscopic order field $\hat x^\alpha({\bf r})$, given for instance by the displacement vector in crystals \cite{SE93,S97,MG20,MG21,H22,MGHF22}.  Its gradient $\hat{u}^{a\alpha}\equiv\nabla^a\hat{x}^\alpha$ obeys a local conservation equation such as Eq.~\eqref{eqs-c} and may thus be considered on the same footing as the microscopic densities of the fundamentally conserved quantities.  
In crystals with a single particle species, the local-equilibrium statistical operator~\eqref{rho-leq} can therefore be defined with the following microscopic densities,
\be
\hat{c}^\alpha = (\hat{e},\hat{g}^a,\hat{\rho},\hat{u}^{a\alpha}) \, ,
\label{broken-c}
\ee
where $\hat{\rho}$ is the mass density and $\hat{u}^{a\alpha}$ are the gradients of the order fields, and with the conjugate fields $\lambda^\alpha=(\lambda_e,\lambda_{g^a},\lambda_{\rho},\lambda_{u^{a\alpha}})$.  More details are given in App.~\ref{AppC}.  

As for fluids, the quantum-mechanical calculation of the transport properties can be achieved in parallel to the classical calculation.  As a consequence of continuous symmetry breaking, the transport properties include further coefficients resulting from the emerging order fields and the anisotropy of the medium.  The transport coefficients are now given by the quantum version~\eqref{eq:GK_formula} of the Green-Kubo formulas.  In centrosymmetric phases, the matrix of transport coefficients is symmetric and the Green-Kubo formulas are equivalent to Einstein-Helfand formulas.  Otherwise, transport processes with $\epsilon_\alpha\epsilon_\beta=-1$ could be coupled together, giving antisymmetric matrix elements, which do not contribute to the entropy production rate \cite{MG20,MG21,H22}.

\section{Conclusion}
\label{Sec:Conclusion}

In this paper, we have carried out the quantum-mechanical derivation of the macroscopic hydrodynamic equations in the local-equilibrium approach for phases of matter without and with broken continuous symmetries.  The statistical operator of the local-equilibrium approach is constructed using the microscopic densities and conjugate fields associated with the slow modes in the system.  

The derivation is performed in the Schr\"odinger picture of quantum mechanics.  For this purpose, we have obtained functional identities and, in particular, Eq.~\eqref{eq:rel1stid} deduced from the variation of Eq.~\eqref{eq:SAfirstid} with respect to the conjugate fields.  The functional identity~\eqref{eq:rel1stid} allows us to obtain a general form for the time evolution of the conjugate fields in the Schr\"odinger picture and to show the equivalence of the local-equilibrium approach with projection-operator methods~\cite{KG73,G77,G82}.

A quantum integral fluctuation theorem is established, from which we deduce the nonnegativity of the entropy production, using the Peierls-Bogoliubov inequality.  With the transformation relating the exact to the local-equilibrium statistical operator, the mean current densities can be decomposed into their reversible and dissipative parts.  Next, the dissipative current densities can be expanded in powers of the affinities defined as the gradients of the conjugate fields.  The transport coefficients are thus identified as the linear response coefficients of the dissipative current densities with respect to the affinities.  The transport properties are given by Green-Kubo formulas in terms of the quantum-mechanical microscopic dynamics.  At leading order of the gradient expansion, the time derivative of the entropy is a quadratic form of the affinities based on the symmetric part of the matrix of transport coefficients.  This symmetric part can be expressed by Einstein-Helfand formulas, proving the nonnegativity of the quadratic form.  For quantum systems, the time derivative of the entropy may thus be identified with the entropy production rate of nonequilibrium thermodynamics \cite{GM84}.  

The results are established for systems ruled by quantum mechanics and they apply not only to fluids, but also to condensed matter phases with broken continuous symmetries.


\section*{Acknowledgements}

The Authors thank the Universit\'e Libre de Bruxelles (ULB) and the Fonds de la Recherche Scientifique - FNRS (Belgium) for support in this research.


\appendix

\section{Formalism calculations}
\label{AppA}

In this appendix, we explicitly derive several equations given in the main text.

\paragraph{The functional derivative~\eqref{drho/dlambda}.} For two possibly noncommuting operators  $\hat{A}$ and  $\hat{B}$, we have the identity
\begin{align}
{\rm e}^{x(\hat{A}+\hat{B})}={\rm e}^{x\hat{A}}+\int_{0}^xdx'{\rm e}^{x'(\hat{A}+\hat{B})}\, \hat{B}\, {\rm e}^{(x-x')\hat{A}}\, ,\label{eq:R66}
\end{align}
which is given for example in Ref.~\cite{R66}. The identity can be proved by considering the derivative of ${\rm e}^{x(\hat{A}+\hat{B})}{\rm e}^{-x\hat{A}}$ with respect to~$x$. From Eq.~\eqref{eq:R66}, the variation of the operator ${\rm e}^{x\hat{A}}$ can be written as
\begin{align}
\delta\, {\rm e}^{x\hat{A}} & \equiv {\rm e}^{x(\hat{A}+\delta\hat{A})}- {\rm e}^{x\hat{A}}\notag\\
&=\int_0^xdx'\,  {\rm e}^{x'\hat{A}}\, (\delta\hat{A})\, {\rm e}^{(x-x')\hat{A}}\, .
\end{align}
The variation of the local-equilibrium statistical operator~\eqref{rho-leq} is thus given by
\begin{align}
\delta \hat{\varrho}_{\text{leq},\boldsymbol{\lambda}} &=-\int_{0}^{1}dx\, {\rm e}^{-x \hat\varsigma({\boldsymbol{\lambda}})}\, \delta \hat\varsigma({\boldsymbol{\lambda}})\, {\rm e}^{x \hat\varsigma({\boldsymbol{\lambda}})} \hat{\varrho}_{\text{leq},\boldsymbol{\lambda}} 
\label{eq:drholeq}
\end{align}
with
\begin{align}
&\delta \hat\varsigma({\boldsymbol{\lambda}}) = \delta{\lambda}^\alpha \ast \hat{c}^\alpha+ \delta \Omega(\boldsymbol{\lambda}) =  \delta{\lambda}^\alpha \ast \delta\hat{c}^\alpha\, ,\label{eq:drho}
\end{align}
using  Eq.~\eqref{eq:dCdlambda} and $\delta\hat{c}^\alpha \equiv \hat{c}^\alpha-\langle\hat{c}^\alpha\rangle_{\text{leq},\boldsymbol{\lambda}}$, whereupon we obtain Eq.~\eqref{drho/dlambda}.
As a consequence, for any operator $\hat{X}$, we have
\begin{align}
& \delta\langle\hat{X}\rangle_{\text{leq},\boldsymbol{\lambda}} = \text{tr}\left[\hat{X} (\delta\hat{\varrho}_{\text{leq},\boldsymbol{\lambda}})\right]\notag\\
&=- \text{tr}\ \hat{X}\int_{0}^{1}dx \, {\rm e}^{-x \hat\varsigma({\boldsymbol{\lambda}})}\, \delta\lambda^\alpha \ast \delta\hat{c}^\alpha \, {\rm e}^{x \hat\varsigma({\boldsymbol{\lambda}})} \hat{\varrho}_{\text{leq},\boldsymbol{\lambda}_t}\notag\\
&=-\int d\mathbf{r} \, \langle \delta \hat{X}^{\dagger}|\delta\hat{c}^\alpha(\mathbf{r}) \rangle_{\boldsymbol{\lambda}} \, \delta \lambda^\alpha(\mathbf{r})\, , \label{eq:varA}
\end{align}
using $\langle\delta\hat{c}^\alpha \rangle_{\text{leq},\boldsymbol{\lambda}} =0$ to replace $ \hat{X}$ by $\delta \hat{X}\equiv \hat{X} - \langle\hat{X}\rangle_{\text{leq},\boldsymbol{\lambda}}$ and the Mori product~\eqref{eq:Moriproduct}, so that
\begin{align}
& \frac{\delta\langle\hat{X}\rangle_{\text{leq},\boldsymbol{\lambda}}}{\delta \lambda^\alpha(\mathbf{r})} = -\langle \delta \hat{X}^{\dagger}|\delta\hat{c}^\alpha(\mathbf{r}) \rangle_{\boldsymbol{\lambda}} = -\langle \delta\hat{c}^\alpha(\mathbf{r})\vert \delta \hat{X} \rangle_{\boldsymbol{\lambda}} \, ,
\label{dXdlambda}
\end{align}
because $\delta\hat{c}^\alpha=\delta\hat{c}^{\alpha\dagger}$.

\paragraph{The functional identity~\eqref{eq:SAfirstid}.} The result follows from $(d/dt){\rm tr}\, \hat\varrho_t|_{t=0}=0$ and
\begin{align}
\frac{d}{dt}{\rm tr}\, \left.\hat\varrho_t\right|_{t=0} 
&=\frac{d}{dt}{\rm tr}\, {\rm e}^{\widehat{\cal L} t} \,\hat{\varrho}_{\text{leq},\boldsymbol{\lambda}_0}\Big|_{t=0} \notag\\
&=\frac{d}{dt}{\rm tr}\, \exp\left[-{\lambda}_0^\alpha\ast \hat{c}_{-t}^\alpha-\Omega(\boldsymbol{\lambda}_0)\right]\Big|_{t=0} \notag\\
&=-{\rm tr}\, {\lambda}_0^\alpha\ast \partial_t\hat{c}_{-t}^\alpha \, \hat\varrho_t\Big|_{t=0}\notag\\
&={\rm tr}\, {\lambda}_0^\alpha\ast \nabla^a\hat{J}^a_{c^\alpha,-t}\, \hat\varrho_t\Big|_{t=0}\notag\\
&=- \nabla^a{\lambda}_0^\alpha \ast \langle \hat{J}^a_{c^\alpha}\rangle_{{\rm leq},{\boldsymbol{\lambda}}_0}\, ,
\label{eq:SAfirstid-proof}
\end{align}
using $\hat{\varrho}_0=\hat{\varrho}_{\text{leq},\boldsymbol{\lambda}_0}$ and setting $\boldsymbol{\lambda}=\boldsymbol{\lambda}_0$.

\paragraph{The functional identity~\eqref{eq:rel1stid}.} 
Since the identity~\eqref{eq:SAfirstid} can be expressed as
 \begin{align}
{\lambda}^\alpha \ast {\nabla}^a\langle\hat{J}^{a}_{c^\alpha}\rangle_{\text{leq},\boldsymbol{\lambda}} &= 0\, ,
\end{align}
its variation with respect to ${\lambda}^\alpha$ gives
\begin{widetext}
\begin{align}
\label{eq:var1idlambda}
\frac{\delta}{\delta \lambda^\alpha(\mathbf{r})}\left[\int d\mathbf{r}'\,  {\lambda}^\beta(\mathbf{r}') \, {\nabla}^a\langle\hat{J}^{a}_{c^\beta}(\mathbf{r}')\rangle_{\text{leq},\boldsymbol{\lambda}} \right]&={\nabla}^a\langle\hat{J}^{a}_{c^\alpha}(\mathbf{r})\rangle_{\text{leq},\boldsymbol{\lambda}} -\int d\mathbf{r}'\, {\lambda}^\beta(\mathbf{r}') \, {\nabla}'^a\langle \delta \hat{c}^{\alpha}(\mathbf{r})| \delta \hat{J}^a_{c^{\beta}}(\mathbf{r}') \rangle_{\boldsymbol{\lambda}}=0\, ,
\end{align}
where we have used
\begin{align}
\delta\left({\lambda}^\beta\ast {\nabla}^a\langle\hat{J}^{a}_{c^\beta}\rangle_{\text{leq},\boldsymbol{\lambda}} \right) &=\delta{\lambda}^\beta\ast {\nabla}^a\langle\hat{J}^{a}_{c^\beta}\rangle_{\text{leq},\boldsymbol{\lambda}} +{\lambda}^\beta\ast {\nabla}^a\delta\langle\hat{J}^{a}_{c^\beta}\rangle_{\text{leq},\boldsymbol{\lambda}}
\end{align}
and Eq.~\eqref{dXdlambda} with $\hat{X}=\hat{J}_{c^\beta}^a$.  Hence, we find the functional identity~\eqref{eq:rel1stid}.

\paragraph{The operator~\eqref{Sigma}.} From its definition, we get
 \begin{align}
\hat{\Sigma}_t(\Gamma) 
& =  \int_0^t {\rm d}\tau\, \left[\partial_\tau {\lambda}_\tau^\alpha \ast \hat{c}^\alpha_{\tau - t} + {\lambda}_\tau^\alpha\ast\partial_\tau \hat{c}^\alpha_{\tau - t} + \partial_\tau\lambda^\alpha_\tau \ast \frac{\delta\Omega(\boldsymbol{\lambda}_\tau)}{\delta \lambda^\alpha_\tau}\right] \notag\\
&= \int_0^t {\rm d}\tau\, \left[ \partial_\tau {\lambda}_\tau^\alpha \ast \left( \hat{c}^\alpha_{\tau - t}-\langle \hat{c}^\alpha\rangle_{{\rm leq},{\boldsymbol{\lambda}}_\tau}\right) - {\lambda}_\tau^\alpha\ast\nabla^a\hat{J}^a_{c^\alpha,{\tau-t}}  \right]\notag\\
&= \int_0^t {\rm d}\tau\, \left[ \partial_\tau {\lambda}_\tau^\alpha \ast \left( \hat{c}^\alpha_{\tau - t}-\langle \hat{c}^\alpha\rangle_{{\rm leq},{\boldsymbol{\lambda}}_\tau}\right) + \nabla^a{\lambda}_\tau^\alpha \ast\left(\hat{J}^a_{c^\alpha,{\tau-t}}-\langle \hat{J}^a_{c^\alpha}\rangle_{{\rm leq},{\boldsymbol{\lambda}}_\tau}\right)\right] ,
\label{Sigma-proof}
\end{align}
\end{widetext}
using Eqs.~\eqref{eqs-c}, \eqref{eq:dCdlambda}, and the first identity~\eqref{eq:SAfirstid} to include $\langle \hat{J}^a_{c^\alpha}\rangle_{{\rm leq},{\boldsymbol{\lambda}}_\tau}$.  We thus obtain the result~\eqref{Sigma2} with the definitions \eqref{delta-c} and \eqref{delta-J}.

\paragraph{The mean value~\eqref{Sigma-DS} of the operator $\hat\Sigma_t$.} Since $\hat\Sigma_t=\ln(\hat{\varrho}_t/\hat{\varrho}_{\text{leq},\boldsymbol{\lambda}_t})$ according to Eqs.~\eqref{rho-leq} and \eqref{rho-Sigma}, we have that
\begin{align}
\langle\hat\Sigma_t\rangle_t 
&= \text{tr} \left(\hat{\varrho}_t\ln \hat{\varrho}_t\right) - \text{tr} \left(\hat{\varrho}_t\ln \hat{\varrho}_{\text{leq},\boldsymbol{\lambda}_t}\right)\notag\\
&=\text{tr} \left(\hat{\varrho}_t\ln \hat{\varrho}_t\right) - \text{tr} \left(\hat{\varrho}_{\text{leq},\boldsymbol{\lambda}_t}\ln \hat{\varrho}_{\text{leq},\boldsymbol{\lambda}_t}\right)\notag\\
&=\frac{1}{k_{\rm B}}\left(-S_0+S_t\right) ,
\label{mean-Sigma-App}
\end{align}
because of the requirements~\eqref{basic_conditions}, the definition~\eqref{entropy} for the entropy $S_t=-k_{\rm B} \text{tr} \left(\hat{\varrho}_{\text{leq},\boldsymbol{\lambda}_t}\ln \hat{\varrho}_{\text{leq},\boldsymbol{\lambda}_t}\right)$, and the von~Neumann entropy $S_0=-k_{\rm B}\text{tr} \left(\hat{\varrho}_t\ln \hat{\varrho}_t\right)$, which is invariant under time evolution.  Defining the relative entropy between two statistical operators $\hat{\varrho}$ and $\hat{\varrho}'$ as
\begin{align}
D(\hat{\varrho}\Vert\hat{\varrho}')\equiv \text{tr} \left(\hat{\varrho}\ln \hat{\varrho}\right) - \text{tr} \left(\hat{\varrho}\ln \hat{\varrho}'\right) \ge 0 \, ,
\end{align}
we find that
\begin{align}
\langle\hat\Sigma_t\rangle_t&=D(\hat{\varrho}_t\Vert\hat{\varrho}_{\text{leq},\boldsymbol{\lambda}_t}) \ge 0 \,.
\end{align}

\paragraph{The operator~\eqref{eq:Xi}.} Setting $\hat{A}=-\hat\varsigma({\boldsymbol{\lambda}}_t)$, $\hat{B}=\hat{\Sigma}_t$, and $x=1$ in Eq.~\eqref{eq:R66}, we obtain
\begin{align}
\hat{\varrho}_t & = {\rm e}^{-\hat\varsigma({\boldsymbol{\lambda}}_t)}+\int_0^1dx\, {\rm e}^{x\left[-\hat\varsigma({\boldsymbol{\lambda}}_t)+\hat{\Sigma}_t\right]} \, \hat{\Sigma}_t \, {\rm e}^{x\hat\varsigma({\boldsymbol{\lambda}}_t)}\, {\rm e}^{-\hat\varsigma({\boldsymbol{\lambda}}_t)}\notag\\
&= \hat{\Xi}_t\, \hat{\varrho}_{\text{leq},\boldsymbol{\lambda}_t}
\label{rho-Xi-leq-App}
\end{align}
with the local-equilibrium statistical operator~\eqref{rho-leq} and the operator $\hat{\Xi}_t$ defined by Eq.~\eqref{eq:Xi}.

\paragraph{The inequality~\eqref{QJI}.} The Peierls-Bogoliubov inequality is given by Eq.~(1.43) of Ref.~\cite{W78}:
\be
{\rm tr} \, {\rm e}^{\hat A + \hat B} \ge {\rm tr} \, {\rm e}^{\hat A + \langle\hat B\rangle} 
\qquad\mbox{with}\qquad
\langle\hat B\rangle = \frac{{\rm tr} \, \hat B \, {\rm e}^{\hat A}}{{\rm tr} \, {\rm e}^{\hat A}} \, .
\label{PB-ineq}
\ee
Taking
\be
\hat A + \hat B = -\hat\varsigma({\boldsymbol{\lambda}}_t) \, , \quad
\hat A = -\hat\varsigma({\boldsymbol{\lambda}}_t)+\hat\Sigma_t \, ,
\quad\mbox{and}\quad
\hat B = - \hat\Sigma_t \, ,
\ee
writing
\be
{\rm tr} \, {\rm e}^{\hat A + \hat B} = {\rm tr} \, {\rm e}^{\hat A} \, {\rm e}^{\hat A + \hat B}  \, {\rm e}^{-\hat A} 
\ee
with the cyclic property of the trace, and using ${\rm tr} \, {\rm e}^{\hat A + \langle\hat B\rangle}={\rm e}^{\langle\hat B\rangle}\,{\rm tr} \, {\rm e}^{\hat A}$ and the normalization condition ${\rm tr} \, {\rm e}^{\hat A}={\rm tr} \, {\rm e}^{-\hat\varsigma({\boldsymbol{\lambda}}_t)+\hat\Sigma_t}={\rm tr} \, \hat\varrho_t=1$, the Peierls-Bogoliubov inequality implies that
\be
{\rm tr} \, {\rm e}^{-\hat\varsigma({\boldsymbol{\lambda}}_t)+\hat\Sigma_t}\, {\rm e}^{-\hat\varsigma({\boldsymbol{\lambda}}_t)}\,  {\rm e}^{\hat\varsigma({\boldsymbol{\lambda}}_t)-\hat\Sigma_t} \ge {\rm e}^{-\langle\hat\Sigma_t\rangle_t}
\label{pre-QJI}
\ee
or, equivalently, the expression~\eqref{QJI}, which is the quantum version of classical Jensen's inequality $\left\langle{\rm e}^{x}\right\rangle \ge {\rm e}^{\langle x\rangle}$ with $x=-\Sigma_t({\boldsymbol{\Gamma}})$.

\paragraph{The entropy time derivative~\eqref{eq:epr}.}  Starting from the entropy~\eqref{eq.entropyfunctional}, we obtain

\begin{align}
\frac{1}{k_{\rm B}}\, \frac{dS}{dt} &= \frac{d}{dt} \left[\lambda_t^\alpha\ast\langle \hat{c}^\alpha\rangle_{{\rm leq},{\boldsymbol{\lambda}}_t}+\Omega(\boldsymbol{\lambda}_t)\right] \notag\\
&=\partial_t\lambda_t^\alpha \ast \langle \hat{c}^\alpha \rangle_{{\rm leq},{\boldsymbol{\lambda}}_t}+\lambda_t^\alpha\ast\partial_t\langle \hat{c}^\alpha\rangle_{t}+\partial_t\lambda_t^\alpha\ast\frac{\delta \Omega(\boldsymbol{\lambda}_t)}{\delta \lambda^\alpha_t} \notag\\
&= - \lambda^{\alpha}_t\ast  \nabla^a\langle \hat{J}^a_{c^{\alpha}}\rangle_{t} = \nabla^a\lambda^{\alpha}_t\ast  \langle \hat{J}^a_{c^{\alpha}}\rangle_{t} \, ,
\label{dS/dt-proof}
 \end{align}
using $\langle \hat{c}^\alpha \rangle_{{\rm leq},{\boldsymbol{\lambda}}_t}=\langle \hat{c}^\alpha\rangle_{t}$ and  Eq.~\eqref{eq:dCdlambda}.

\begin{widetext}

\paragraph{Equation~\eqref{eq:dlambdadt} for the conjugate fields.} With the requirements $\langle\hat{c}^{\alpha}(\mathbf{r})\rangle_t= \langle\hat{c}^{\alpha}(\mathbf{r})\rangle_{\text{leq},\boldsymbol{\lambda}_t}$ and Eq.~\eqref{eq:varA} with $\delta$ replaced by $\partial_t$ and $\hat{X}$ by $\hat{c}^\alpha$, the time derivatives of the mean densities can be expressed as
\begin{align}
&\partial_t  \langle\hat{c}^{\alpha}(\mathbf{r})\rangle_{\text{leq},\boldsymbol{\lambda}_t} = \text{tr}\left[ (\partial_t\hat{\varrho}_{\text{leq},\boldsymbol{\lambda}_t})\, \hat{c}^{\alpha}(\mathbf{r})\right] = -\int d\mathbf{r}' \, \langle \delta \hat{c}^{\alpha}(\mathbf{r})|\delta\hat{c}^\beta(\mathbf{r}') \rangle_{\boldsymbol{\lambda}_t} \, \partial_t\lambda_t^\beta(\mathbf{r}') \, . \label{eq:dtcalpha}
\end{align}
Multiplying by $\langle \delta \hat{c}^{\gamma}(\mathbf{r''})|\delta\hat{c}^\alpha(\mathbf{r}) \rangle^{-1}_{\boldsymbol{\lambda}_t}$ on the left, integrating over $d\mathbf{r}$, and using Eq.~\eqref{eqs-macro-c} give the result:
\begin{align}
\partial_t\lambda_t^\alpha(\mathbf{r})&=-\int d\mathbf{r}' \, \langle \delta \hat{c}^{\alpha}(\mathbf{r})|\delta\hat{c}^\beta(\mathbf{r}') \rangle^{-1}_{\boldsymbol{\lambda}_t} \, \partial_t \langle\hat{c}^{\beta}(\mathbf{r}')\rangle_{\text{leq},\boldsymbol{\lambda}_t} \notag\\ 
&=-\int d\mathbf{r}' \, \langle \delta \hat{c}^{\alpha}(\mathbf{r})|\delta\hat{c}^\beta(\mathbf{r}') \rangle^{-1}_{\boldsymbol{\lambda}_t} \, \partial_t  \langle\hat{c}^{\beta}(\mathbf{r}')\rangle_t =\int d\mathbf{r}' \, \langle \delta \hat{c}^{\alpha}(\mathbf{r})|\delta\hat{c}^\beta(\mathbf{r}') \rangle^{-1}_{\boldsymbol{\lambda}_t} \, \nabla'^a \langle \hat J^a_{c^\beta}({\bf r}')\rangle_t \, .
\end{align}

\paragraph{Equation~\eqref{eq:decfprop} introducing the projection operator.}  Using the decomposition~\eqref{mean_crnt_dens}, we then have
\begin{align}
\partial_t {\lambda}_t^\alpha \ast\delta\hat{c}^\alpha &=\iint d\mathbf{r}\, d\mathbf{r}' \, \delta\hat{c}^\alpha(\mathbf{r}) \, \langle \delta \hat{c}^{\alpha}(\mathbf{r})|\delta\hat{c}^\beta(\mathbf{r}') \rangle^{-1}_{\boldsymbol{\lambda}_t} \, \nabla'^a \langle \hat J^a_{c^\beta}({\bf r}')\rangle_t \notag\\
&=\iint d\mathbf{r}\, d\mathbf{r}' \, \delta\hat{c}^\alpha(\mathbf{r}) \, \langle \delta \hat{c}^{\alpha}(\mathbf{r})|\delta\hat{c}^\beta(\mathbf{r}') \rangle^{-1}_{\boldsymbol{\lambda}_t} \left[ \nabla'^a \bar{J}^a_{c^\beta}({\bf r}',t)+\nabla'^a  \mathcal{J}^a_{c^\beta}({\bf r}',t)\right] .\label{eq:partlambadeltac}
\end{align}
Next, the functional identity~\eqref{eq:rel1stid} gives Eq.~\eqref{eq:decf}, which becomes
\begin{align}
\partial_\tau {\lambda}_\tau^\alpha \ast\delta\hat{c}^\alpha &=-\iiint d\mathbf{r}\, d\mathbf{r}' \, d\mathbf{r}'' \, \delta\hat{c}^\alpha(\mathbf{r}) \, \langle \delta \hat{c}^{\alpha}(\mathbf{r})|\delta\hat{c}^\beta(\mathbf{r}') \rangle^{-1}_{\boldsymbol{\lambda}_\tau} \, \langle \delta \hat{c}^{\beta}(\mathbf{r}')| \delta \hat{J}^a_{c^{\gamma}}(\mathbf{r}'') \rangle_{\boldsymbol{\lambda}_\tau} \, \nabla''^a{\lambda_\tau^\gamma}(\mathbf{r}'')+O(\nabla^2)\notag\\
&=-\iiint d\mathbf{r}\, d\mathbf{r}' \, d\mathbf{r}'' \, \delta\hat{c}^\alpha(\mathbf{r}) \, \frac{\delta \lambda^\beta(\mathbf{r}')}{\delta c^\alpha(\mathbf{r})} \, \frac{\delta \langle \hat{J}^a_{c^{\gamma}}(\mathbf{r}'') \rangle_{\boldsymbol{\lambda}_\tau}}{\delta \lambda^\beta(\mathbf{r}')} \, \nabla''^a{\lambda_\tau^\gamma}(\mathbf{r}'')+O(\nabla^2)\notag\\
&=- \delta\hat{c}^\alpha \ast \frac{\delta \langle \hat{J}^a_{c^{\beta}} \rangle_{\boldsymbol{\lambda}_\tau}}{\delta c^\alpha}  \ast \nabla^a{\lambda_\tau^\beta} +O(\nabla^2) = - \nabla^a{\lambda_\tau^\alpha} \ast\widehat{\mathcal{P}}_{\boldsymbol{\lambda}_\tau}\delta \hat{J}^a_{c^{\alpha}}  +O(\nabla^2)\, ,
\label{eq:dlambda-projection}
\end{align}
\end{widetext}
using Eqs.~\eqref{dlambda/dc}, \eqref{dXdlambda} with $\hat{X}=\hat{J}_{c^\gamma}^a$, and, finally, the projector operator $\widehat{\mathcal{P}}_{\boldsymbol{\lambda}}$ defined in Eq.~\eqref{eq:pop}.
We note that the property $\widehat{\mathcal{P}}^2_{\boldsymbol{\lambda}}\hat{X}=\widehat{\mathcal{P}}_{\boldsymbol{\lambda}}\hat{X}$ can be inferred from the definition~\eqref{eq:pop} of $\widehat{\mathcal{P}}_{\boldsymbol{\lambda}}$.

\paragraph{The Onsager-Casimir reciprocal relations~\eqref{OCRR}.} 
The time-dependent response functions defined as
\be
\phi^{ab}_{\alpha\beta}(t) \equiv \frac{k_{\rm B}T}{V} \int_0^{(k_{\rm B}T)^{-1}} \left\langle \delta\hat{ \mathbb J}_{c^\alpha}^{\prime a}(t) \, \delta\hat{{\mathbb J}}^{\prime b}_{c^\beta}(\imath\hbar\vartheta)\right\rangle_{\rm eq} d\vartheta 
\label{phi(t)}
\ee
have the following general property
\be
\phi^{ab}_{\alpha\beta}(t) = \phi^{ba}_{\beta\alpha}(-t) \, ,
\label{phi-symm}
\ee
which is proved with the change of integration variable $\vartheta'=(k_{\rm B}T)^{-1}-\vartheta$ \cite{K57}.

Moreover, the microdynamics is symmetric under the antiunitary time-reversal transformation $\hat\Theta$ such that
\be
\hat\Theta\, \imath\, \hat\Theta^{-1} = -\imath \, , \quad \hat\Theta\, \hat H\, \hat\Theta^{-1} = \hat H \, , \quad \hat\Theta \,\delta\hat{\mathbb J}^{\prime a}_{c^\alpha} \hat\Theta^{-1} = \epsilon_{\alpha} \, \delta\hat{\mathbb J}^{\prime a}_{c^\alpha}
\ee
with $\epsilon_{\alpha}=\pm 1$.  Consequently, the response functions~\eqref{phi(t)} satisfy
\be
\phi^{ab}_{\alpha\beta}(t) = \epsilon_{\alpha}\, \epsilon_{\beta} \, \phi^{ab}_{\alpha\beta}(-t) \, .
\label{TR-phi}
\ee
Since the linear response coefficients are given by integrating the response functions over time according to the Green-Kubo formulas~\eqref{eq:GK_formula}
\be
{\cal L}^{ab}_{\alpha\beta} = \int_0^{\infty} \phi^{ab}_{\alpha\beta}(t) \, dt \, ,
\label{L-int-phi}
\ee
Eq.~\eqref{TR-phi} combined with Eq.~\eqref{phi-symm} implies the Onsager-Casimir reciprocal relations~\eqref{OCRR}.

\paragraph{Einstein-Helfand formulas~\eqref{EH-formula}.}  For the response functions~\eqref{phi(t)}, we have the following identity,
\be
\int_0^t dt' \int_0^t dt'' \, \phi^{ab}_{\alpha\beta}(t'-t'') = \int_{-t}^{+t} d\tau \left( t-\vert\tau\vert\right) \phi^{ab}_{\alpha\beta}(\tau) \, ,
\ee
which is obtained with the changes of integration variables $\tau=t'-t''$ and ${\cal T}=(t'+t'')/2$, such that $dt'dt''=d\tau d{\cal T}$.  In the limit $t\to\infty$ and using Eq.~\eqref{phi-symm}, we thus have
\bea
&& \lim_{t\to\infty} \frac{1}{2t} \int_0^t dt' \int_0^t dt'' \, \phi^{ab}_{\alpha\beta}(t'-t'') \nonumber\\
&& = \frac{1}{2} \int_{-\infty}^{+\infty} \phi^{ab}_{\alpha\beta}(t) \, dt = {\mathcal L}^{ab\,{\rm S}}_{\alpha\beta} \, ,
\label{XXX}
\eea
giving the symmetrized transport coefficients~\eqref{symm-L}. Inserting the expressions~\eqref{phi(t)} for the response functions, we obtain the Einstein-Helfand formulas~\eqref{EH-formula} for the symmetrized transport coefficients
in terms of the quantum-mechanical versions~\eqref{Helfand-H} of the Helfand moments, which have the following more explicit forms,
\be\label{Helfand-H-2}
\hat{\mathbb H}_{c^\alpha}^{\prime a}(t,\vartheta) = \int_{0}^{t} {\rm e}^{\frac{\imath}{\hbar}\hat H \tau} \, {\rm e}^{-\frac{\vartheta}{2}\hat H} \, \delta\hat{\mathbb J}_{c^\alpha}^{\prime a} \, {\rm e}^{\frac{\vartheta}{2}\hat H} \, {\rm e}^{-\frac{\imath}{\hbar}\hat H \tau} \, d\tau \, ,
\ee
so that the adjoint operators are given by
\be\label{Helfand-H-2-adj}
\hat{\mathbb H}_{c^\alpha}^{\prime a\dagger}(t,\vartheta) = \int_{0}^{t} {\rm e}^{\frac{\imath}{\hbar}\hat H \tau} \, {\rm e}^{\frac{\vartheta}{2}\hat H} \, \delta\hat{\mathbb J}_{c^\alpha}^{\prime a} \, {\rm e}^{-\frac{\vartheta}{2}\hat H} \, {\rm e}^{-\frac{\imath}{\hbar}\hat H \tau} \, d\tau \, ,
\ee
since the operators $\delta\hat{\mathbb J}_{c^\alpha}^{\prime a}$ are self-adjoint.

\vspace{5mm}

\section{Multicomponent fluids}
\label{AppB}

\setcounter{paragraph}{0}

\paragraph{Hamiltonian dynamics.}
We consider a nonreactive multicomponent fluid, where all the hydrodynamic modes originate from the fundamental conservation laws. 

We adopt the following notations: $N$ is the total number of particles. $\nu$ is the total number of species. The indices $i,j,\ldots= 1, 2,\dots, N$ label the particles and $k,l,\ldots = 1, 2, \dots, \nu$ the particle species. The indices for the particles of species $k$ are denoted $i\in{\cal S}_k$.

The slow modes of the system are the energy density, the linear momentum density, and the particle densities of each species. In compact notations, they are denoted as $\hat{c}^\alpha = (\hat{e},\hat{g}^b,\hat{n}_k) $ and the corresponding current densities are $\hat{J}_{c^\alpha}^a = (\hat{J}^a_{e},\hat{J}^a_{ g^b},\hat{J}^a_{n_k})$.

The Hamiltonian operator is 
\be\label{Hamiltonian-ineq}
\hat H = \sum_{i=1}^N \frac{\hat{\bf p}_i^2}{2m_i} + \sum_{1\leq i<j\leq N} u_{ij} \, , 
\ee
where, in the position representation, $\hat{\bf p}_i=-\imath\hbar\boldsymbol{\nabla}_i$ is the linear momentum of the $i^{\rm th}$ particle, ${\bf r}_i\in{\mathbb R}^3$ its position, and $u_{ij}=u_{ij}(r_{ij})$ with $r_{ij}=\Vert{\bf r}_i-{\bf r}_j\Vert$ the potential energy of the binary interaction between the particles $i$ and $j$.

\paragraph{Microscopic densities and current densities.} The particle, mass, momentum, and energy microscopic densities can be defined as follows in the quantum formulation,
\bea
\hat n_k({\bf r}) &=& \sum_{i\in{\cal S}_k} \delta ({\bf r} - {\bf r}_{i}) \, , \label{k-particle_dens}\\
\hat \rho({\bf r}) &=& \sum_{i=1}^N m_i \, \delta ({\bf r} - {\bf r}_{i}) = \sum_{k=1}^{\nu} m_k \, \hat n_k({\bf r}) \, , \label{mass_dens}\\
\hat{g}^a({\bf r}) &=& \frac{1}{2} \sum_{i=1}^{N} \left[ \hat {p}^a_{i} \,\delta ({\bf r} - {\bf r}_{i}) + \delta ({\bf r} - {\bf r}_{i}) \, \hat {p}^a_{i} \right] , \label{mom_dens}\\
\hat{e}({\bf r}) &=& \frac{1}{2} \sum_{i=1}^{N} \left[ \hat\varepsilon_{i}\, \delta ({\bf r}-{\bf r}_{i}) + \delta ({\bf r}-{\bf r}_{i}) \, \hat\varepsilon_{i}\right] , \label{energy_density}
\eea
where $\hat\varepsilon_{i}$ is the Hermitian operator giving the energy of the $i^{\rm th}$ particle according to
\be\label{E_a}
\hat\varepsilon_{i} = \frac{\hat{\bf p}_i^2}{2m_i} + \frac{1}{2} \sum_{\scriptstyle j=1 \atop\scriptstyle (j\neq i)}^{N}  u_{ij} \, .
\ee
The current densities obtained from Eq.~\eqref{eqs-c} read
\begin{widetext}
\begin{align}
\hat{J}^a_{n_k}({\bf r}) &= \frac{1}{2} \sum_{i\in{\cal S}_k} \left[ \frac{\hat{p}^a_i}{m_i} \, \delta({\bf r}-{\bf r}_{i}) + \delta({\bf r}-{\bf r}_{i}) \, \frac{\hat{p}_i^a}{m_i}\right] , \label{k-particle-crnt_dens}
\end{align}
\begin{align}
\hat{J}^a_{g^b}({\bf r}) &=\sum_{i} \frac{1}{4m_i} \left[ \hat p_i^{a}\, \hat p_i^{b}\, \delta({\bf r}-{\bf r}_{i}) + \hat p_i^{a}\, \delta({\bf r}-{\bf r}_{i}) \, \hat p_i^{b} + \hat p_i^{b}\, \delta({\bf r}-{\bf r}_{i}) \,  \hat p_i^{a}+ \delta({\bf r}-{\bf r}_{i}) \, \hat p_i^{b}\, \hat p_i^{a}\right] + \frac{1}{2} \sum_{i\neq j} F_{ij}^{b} D_{ij}^a({\bf r}) \, ,\label{micro-pressure}
\end{align}
\begin{align}
\hat{J}^a_{e}({\bf r}) &= \frac{1}{4} \sum_{i} \hat \varepsilon_{i}\left[ \frac{\hat{p}^a_i}{m_i} \, \delta({\bf r}-{\bf r}_{i}) + \delta({\bf r}-{\bf r}_{i}) \, \frac{\hat{p}_i^a}{m_i}\right] + \frac{1}{16} \sum_{i\neq j} \left[\left(\frac{\hat{p}^b_{i}}{m_i}+\frac{\hat{p}^b_{j}}{m_j}\right) {F}^{b}_{ij} + {F}^{b}_{ij}\left(\frac{\hat{p}^b_{i}}{m_i}+\frac{\hat{p}^b_{j}}{m_j}\right)\right] D_{ij}^a({\bf r})  + \mbox{H. c.}\, , \label{micro_dens_je}
\end{align}
\end{widetext}
with
\begin{align}
D_{ij}^a({\bf r}) \equiv  \int_0^1 d\xi \, \frac{d{R}^a_{ij}(\xi)}{d\xi}\, \delta[ {\bf r}-\boldsymbol{R}_{ij}(\xi)] \, ,
\label{Daij}
\end{align}
where $F^a_{ij}\equiv -(\partial/\partial r^a_{i}) u_{ij}(r_{ij})$ is the force exerted on the particle~$i$ by the particle~$j$ and $\boldsymbol{R}_{ij}(\xi)$ is a smooth curve joining $\boldsymbol{R}_{ij}(0)={\bf r}_{j}$ to $\boldsymbol{R}_{ij}(1)={\bf r}_{i}$, e.g., $\boldsymbol{R}_{ij}(\xi)={\bf r}_j+\xi({\bf r}_i-{\bf r}_j)$, in which case $d{R}^a_{ij}/d\xi=r_{ij}^a\equiv r_i^a-r_j^a$.

\paragraph{Local thermodynamics.} In the laboratory frame, where the continuous medium moves with the velocity ${\bf v}=(v^a)$, the energy density and the chemical potential of species $k$ are respectively given by
\be\label{eps-mu}
e=e_0+\rho{\bf v}^2/2 \qquad\mbox{and} \qquad \mu_k=\mu_{k0}-m_k{\bf v}^2/2
\ee
in terms of the corresponding quantities $e_0$ and $\mu_{k0}$ in the moving frame, where the medium is at rest.  Therefore, in the laboratory frame, the Euler and Gibbs relations have the following forms up to terms of $O(\nabla^2)$,
\bea
s &=& \frac{e+p}{T}-\sum_k \frac{\mu_k}{T}\, n_k -\frac{v^a}{T}\, g^a \, , \label{Euler-thermo-rel} \\
ds &=& \frac{1}{T}\, de -\sum_k \frac{\mu_k}{T}\, d n_k -\frac{v^a}{T}\, dg^a \, , \label{s-Euler+Gibbs}
\eea
where $s$ is the entropy density introduced in Eq.~\eqref{S-local-s} and $p$ is the pressure.  From Eq.~\eqref{eq:SAsecondid}, the conjugate fields at leading order in the gradients are given in terms of the inverse temperature $\beta\equiv(k_{\rm B}T)^{-1}$ by
\bea
&&\lambda_{e}({\bf r},t) \equiv \frac{1}{k_{\rm B}} \, \frac{\delta S({\bf c})}{\delta e({\bf r},t)} =\beta({\bf r},t) \, , \label{chi-e}
\\
&&\lambda_{n_k}({\bf r},t) \equiv \frac{1}{k_{\rm B}} \, \frac{\delta S({\bf c})}{\delta n_k({\bf r},t)} =-\beta({\bf r},t)\, \mu_k({\bf r},t) \, , \qquad \label{chi-n_k}
\\
&&\lambda_{g^a}({\bf r},t) \equiv \frac{1}{k_{\rm B}} \, \frac{\delta S({\bf c})}{\delta g^a({\bf r},t)} =-\beta({\bf r},t)\, v^a({\bf r},t)  \, , \label{chi-g_i}
\eea
again up to terms of $O(\nabla^2)$ as in the classical framework.

\paragraph{Reversible current densities.} The reversible current densities are derived by a direct computation of the statistical averages with the changes of momentum variables $\hat{\bf p}_i=\hat{\bf p}_{i0}+m_i{\bf v}({\bf r}_i)$ from the laboratory frame to the frame moving with the medium.  Consequently, the observables of interest are transformed into the corresponding observables with the subscript $0$ according to
\begin{align}
 \hat{e} &= \hat{e}_0 + \hat{g}^a_0{v}^a + \frac{1}{2}\, \hat{\rho} \, {\bf v}^2\, ,\label{e-e0} \\
  \hat{g}^a& = \hat{g}^a_0 + \hat{\rho}\, {v}^a \; ,\label{g-g0}\\
\hat{J}^a_{n_k}&=\hat{J}^a_{n_k,0} + \hat{n}_k\, {v}^a \; ,  \qquad \label{jk-jk0}\\
\hat{J}^a_{g^b}&= \hat{J}^a_{g^b,0} + \hat{g}^a_0\,{v}^b + {v}^a\,\hat{g}^b_0+\hat{\rho}\, {v}^a\, {v}^b \, ,\label{tau-tau0}\\
\hat{J}^a_{e} &=\hat{J}^a_{e,0} + \hat{e}_0\, {v}^a +\hat{J}^a_{g^b,0} {v}^b + \hat{g}^b_0{v}^b\, {v}^a + \frac{1}{2}\, {\bf v}^2 (\hat{g}^a_0+\hat{\rho}\,{v}^a) \notag\\
&\quad - \left(\hat{\Delta}^a + \hat{\Delta}^{\prime a} \right) ,  \label{je-je0}
\end{align}
where
\be
\hat{\Delta}^a \equiv \frac{1}{2} \sum_{i\ne j} \left[ {v}^b({\bf r})-\frac{{v}^b({\bf r}_i)+{v}^b({\bf r}_j)}{2}\right] { F}^b_{ij} \, D_{ij}^a({\bf r})
\ee
expressed with the quantity~\eqref{Daij} \cite{S14}, and
\be
\hat{\Delta}^{\prime a} \equiv \frac{\hbar^2}{4} \, {\nabla}^b \left[ ( {\nabla}^a{ v}^b) \sum_i \frac{1}{m_i} \, \delta({\bf r}-{\bf r}_i) \right] .  
\ee
We evaluate the statistical averages over local equilibrium in the frame moving with the medium, which gives $\langle\hat{g}^a_0\rangle_{\rm leq} =0$, $\langle\hat{J}^a_{n_k,0}\rangle_{\rm leq}=0$, $\langle\hat{J}^a_{e, 0}\rangle_{\rm leq}=0$, $\langle\hat\rho\rangle_{\rm leq}=\rho$, $\langle\hat{e}_0\rangle_{\rm leq}=e_0$, $\langle\hat n_k\rangle_{\rm leq}=n_k$, and $\langle\hat{J}^a_{g_0^b}\rangle_{\rm leq}=p\,\delta^{ab}$.  In addition, $\langle\hat{\Delta}^a+\hat{\Delta}^{\prime a }\rangle_{\rm leq}$ behaves as the square of the gradients and can thus be neglected.  Therefore, we get
\begin{align}
e &=\langle\hat{e}\rangle_{\rm leq} =e_0+\frac{\rho}{2}\,{\bf v}^2 \, ,\label{e-leq}\\
{g}^a&=\langle\hat{g}^a\rangle_{\rm leq}  =\rho\,{v}^a \, ,  \label{g-leq}\\
 \bar{J}^a_{n_k}&=\langle\hat{J}^a_{n_k}\rangle_{\rm leq} =n_k {v}^a\, ,\label{j_nk-leq}\\
\bar{J}^a_{g^b}&=\langle\hat{J}^a_{g^b}\rangle_{\rm leq} = \rho{ v}^a{v}^b + p\, \delta^{ab} \, , \label{j_g-leq}\\ 
\bar{J}^a_{e}& =\langle\hat{J}^a_{e}\rangle_{\rm leq} =\left( e_0+p+\frac{\rho}{2}\,{\bf v}^2\right) {v}^a +O(\nabla^2) \, ,\label{j_e-leq}
\end{align}
for the mean value of the energy density and the reversible parts of the mean current densities.
Introducing these expressions into the macroscopic equations~\eqref{eqs-macro-c} leads to the dissipativeless Eulerian equations of hydrodynamics, which thus satisfy the Galilean invariance.  

\paragraph{Entropy production rate and dissipative current densities.}

Substituting the conjugate fields (\ref{chi-e})-(\ref{chi-g_i}) into the expression~\eqref{eq:entropyrate-dissip_crnt}, the time derivative of the entropy reads
\be
\frac{1}{k_{\rm B}} \frac{dS}{dt} = \int \bigg[{\nabla}^a\beta\, {\cal J}^a_{e} - \nabla^a(\beta{v}^b) {\cal J}^a_{g^b} -\sum_k {\nabla}^a(\beta\mu_k) {\cal J}^a_{n_k} \bigg] d{\bf r} 
\ee
at leading order of the expansion in the gradients.  Using $\nabla^a(\beta v^b)=\beta\nabla^a v^b+v^b\nabla^a\beta$, Eq.~(\ref{eps-mu}) for the chemical potentials, and the relation $\sum_k m_k \,{\cal J}^a_{n_k}=0$ implied by mass conservation, we find
\begin{align}
\frac{1}{k_{\rm B}} \frac{dS}{dt} & = \int \bigg[{\nabla}^a\beta\left({\cal J}^a_{e} -v^b{\cal J}^a_{g^b}\right) - \beta(\nabla^a{v}^b)\, {\cal J}^a_{g^b} \notag\\
&\qquad\qquad\qquad -\sum_k {\nabla}^a(\beta\mu_{k0})\, {\cal J}^a_{n_k} \bigg] d{\bf r}  \, .
\end{align}
The heat current density can thus be identified as
\be\label{J_q-dfn}
{\cal J}^a_q \equiv {\cal J}^a_{e} - {v}^b{\cal J}^a_{g^b}\, ,
\ee
the diffusive current density of species $k$ as ${\cal J}^a_k$, the dissipative part of the pressure tensor as $
{\Pi}^{ab} \equiv {\cal J}^a_{ g^b}$, and the phenomenological affinities mentioned in Eq.~\eqref{ph_Aff} as 
\be\label{Aff_fluid}
{\mathcal A}^a_{c^\alpha,{\rm ph}} = [{\nabla}^a\beta,  - \beta(\nabla^a{v}^b), -{\nabla}^a(\beta\mu_{k0})] \, .
\ee

\paragraph{Hydrodynamic equations.}

Replacing the dissipativeless and dissipative current densities into the balance equations, the equations of fluid mechanics are obtained:
\begin{align}
&\partial_t n_k+\nabla^a(n_k\, v^a+{\cal J}^a_k)=0 \, , \label{dissip-eq-nk}\\
&\partial_t\rho+\nabla^a(\rho\, v^a)=0 \, , \label{dissip-eq-mass}\\
&\partial_t(\rho\, v^a)+\nabla^b\big(\rho\, v^a\, v^b + p \, \delta^{ab} + \Pi^{ab}\big) = 0 \, , \label{dissip-eq-g}\\
&\partial_t\Big(e_0+\frac{\rho}{2}\,{\bf v}^2\Big) + \nabla^a\Big[\Big(e_0+p+\frac{\rho}{2}\,{\bf v}^2\Big) v^a \qquad\nonumber\\
&\qquad\qquad\qquad\qquad + \, \Pi^{ab} \, v^b +{\cal J}^a_q \Big] = 0 \, . \label{dissip-eq-etot}
\end{align}
They have the same form as in the classical framework, as established on the basis of nonequilibrium thermodynamics \cite{GM84}.  In particular, Eq.~(\ref{dissip-eq-g}) is equivalent to the Navier-Stokes equations \index{Navier-Stokes equations} of hydrodynamics.  Therefore, the microscopic approach based on the local-equilibrium statistical operator~(\ref{rho-leq}) and the expansion in powers of the gradients justifies the rules of nonequilibrium thermodynamics for the linear transport properties.

\paragraph{Green-Kubo formulas.}

 The dissipativeless equations for the time evolution of the conjugate fields are given by
 \begin{align}
 \partial_t\,\beta &= -v^a \nabla^a\beta + \beta \, \chi \, \nabla^a v^a \, , \label{dbeta_dt} \\
\partial_t(\beta\mu_{k0}) &=-v^a \nabla^a(\beta\mu_{k0}) - \beta \, \psi_k \, \nabla^a v^a \, , \label{dbetamu_dt} \\
\beta\, \partial_t\, v^a &= -\beta\, v^b \nabla^b \, v^a + \rho^{-1} (e_0+p)\, \nabla^a\beta\nonumber \\
&\qquad\qquad\quad  - \sum_k \rho^{-1} n_k \, \nabla^a (\beta \mu_{k0}) \, ,  \label{dv_dt}
\end{align}
with the coefficients
\begin{align}
\chi &\equiv \left(\frac{\partial p}{\partial e_0}\right)_{\{n_k\}} \, , \\
\psi_k &\equiv \left(\frac{\partial p}{\partial n_k}\right)_{e_0,\{n_j\}_{j\ne k}} \, .
\end{align}
These equations are obtained from Maxwell's thermodynamic relations and the hydrodynamic equations \eqref{dissip-eq-nk}-\eqref{dissip-eq-etot} by neglecting the dissipative terms, since they are of higher order in the gradients.  Accordingly, the operator $\hat{\Sigma}_t$ given by Eq.~\eqref{Sigma} can be expressed as
\begin{align}
\hat{\Sigma}_t & = \int_0^t {\rm d}\tau\,  \bigg[{\nabla}^a\beta_\tau\ast\delta\hat{J}^{\prime a}_{e,{\tau-t}}  
-\beta_\tau{\nabla}^a{v}^b_\tau\ast\delta\hat{J}^{\prime a}_{g^b,\tau-t} \notag\\
&\qquad\qquad\qquad -\sum_k {\nabla}^a(\beta_\tau\mu_{k0,\tau})\ast\delta\hat{J}^{\prime a}_{n_{k},\tau-t}\bigg] 
\end{align}
with
\begin{align}
&\delta\hat{J}^{\prime a}_{e} \equiv \delta\hat{J}^{a}_{e}-\rho^{-1} (e_0+p) \, \delta\hat{g}^a \, , \label{delta_je} \\
&\delta\hat{J}^{\prime a}_{g^b}\equiv \delta\hat{J}^{ a}_{g^b} -\bigg( \chi \,\delta\hat{e} + \sum_k \psi_k \, \delta\hat{n}_k\bigg)\delta^{ab}\, , \label{delta_tau} \\
&\delta\hat{J}^{\prime a}_{n_{k}}\equiv\delta\hat{J}^{ a}_{n_{k}} - \rho^{-1} n_k \, \delta\hat{g}^a \, . \label{delta_jk}
\end{align}
These expressions are here derived from Eqs.~\eqref{dissip-eq-nk}-\eqref{dissip-eq-etot} for the conjugate fields.  We note that they can also be derived using the projection operator~\eqref{eq:pop} and Eq.~\eqref{prime-cd}.
 The microscopic total currents~\eqref{tot_micro_crnt} are thus given by
\begin{align}
&\delta\hat{\mathbb J}^{\prime a}_{e}  =\delta\hat{\mathbb J}^{a}_{e} - \rho^{-1} (e_0+p) \, \delta\hat{P}^a\, , \label{micro-Je} \\
&\delta\hat{\mathbb J}^{\prime a}_{g^b} = \delta\hat{\mathbb J}^{a}_{g^b} -\bigg( \chi \,\delta\hat{E} + \sum_k \psi_k \,  \delta\hat{N}_k\bigg) \delta^{ab}\, , \label{micro-Jg} \\
&\delta\hat{\mathbb J}^{\prime a}_{n_{k}} = \delta\hat{\mathbb J}^{a}_{n_{k}} - \rho^{-1} n_k \, \delta\hat{P}^a \, , \label{micro-Jk}
\end{align}
where $\delta\hat{P}^a=\int \delta\hat{g}^a \, d{\bf r}$, $\delta\hat{E}=\int\delta\hat{e}\, d{\bf r}$, and $\delta\hat{N}_k=\int \delta\hat{n}_k \, d{\bf r}$ are the fluctuations $\delta \hat{C}^\alpha \equiv \hat{C}^\alpha-\langle \hat{C}^\alpha \rangle_{\rm eq}$ with respect to equilibrium for the total momentum, energy, and particle numbers.  Since fluids are isotropic, the fourth-order tensor of viscosity coefficients has the following form,
\be
\eta^{abcd} = \eta\left(\delta^{ac}\delta^{bd}+\delta^{ad}\delta^{bc}-\frac{2}{3}\, \delta^{ab}\delta^{cd}\right)+\zeta \, \delta^{ab}\delta^{cd} \, ,
\ee
and vectorial quantities cannot be coupled to tensorial ones.  Therefore, the viscous part of the pressure tensor, the heat current density, and the diffusive current density of species $k$ are obtained as
\begin{align}
{\cal J}^a_{g^b} &= -\eta \left( {\nabla}^a{v}^b+\nabla^b{v}^a- \frac{2}{3} \, {\nabla}^c{v}^c\delta^{ab}\right)-\zeta\, {\nabla}^c v^c\, \delta^{ab} \, , \label{dissip-Pi}
\\
{\cal J}^a_{q} &= {\cal L}_{qq}\, \nabla^a\frac{1}{T}  + \sum_k {\cal L}_{qk} \, \nabla^a\left(-\frac{\mu_{k0}}{T}\right) , \label{dissip-J_q}
\\
{\cal J}^a_{n_k} &= {\cal L}_{kq}\, \nabla^a\frac{1}{T} + \sum_l {\cal L}_{kl} \, \nabla^a\left(-\frac{\mu_{l0}}{T}\right) .\label{dissip-J_k}
\end{align}
The shear viscosity $\eta$, the bulk viscosity $\zeta$, the heat conductivity $\kappa={\cal L}_{qq}/T^2$, the thermodiffusion coefficients ${\cal L}_{kq}$, and the diffusion coefficients ${\mathcal D}_k=({\mathcal L}_{kk}/T)(\partial\mu_{k0}/\partial n_k)_T$ can thus be evaluated in the limit $V\to\infty$ with the Green-Kubo formulas:
\begin{align}
&\eta = \dfrac{1}{V} \int_0^\infty dt \int_0^{\beta} d\vartheta \, \langle \delta\hat{\mathbb J}'_{xy}(t)\, \delta\hat{\mathbb J}'_{xy}(\imath\hbar\vartheta)\rangle_{\rm eq} \, , \label{GK-eta}
\\
&\zeta = \dfrac{1}{V} \int_0^\infty dt \int_0^{\beta} d\vartheta \, \langle \delta\hat{\mathbb J}'_{\zeta}(t)\, \delta\hat{\mathbb J}'_{\zeta}(\imath\hbar\vartheta)\rangle_{\rm eq} \, , \label{GK-zeta+eta}
\\
&{\cal L}_{qq} = \frac{T}{V} \int_0^{\infty} dt \int_0^{\beta} d\vartheta \, \langle \delta\hat{\mathbb J}'_{e x}(t) \, \delta\hat{\mathbb J}'_{e x}(\imath\hbar\vartheta) \rangle_{\rm eq} \, , \label{GK-L_qq}
\\
&{\cal L}_{kq} = \frac{T}{V} \int_0^{\infty} dt \int_0^{\beta} d\vartheta \, \langle \delta\hat{\mathbb J}'_{kx}(t) \, \delta\hat{\mathbb J}'_{e x}(\imath\hbar\vartheta) \rangle_{\rm eq} \, , \qquad\ \label{GK-L_kq}
\\
&{\cal L}_{kl} = \frac{T}{V} \int_0^{\infty} dt \int_0^{\beta} d\vartheta \, \langle \delta\hat{\mathbb J}'_{kx}(t) \, \delta\hat{\mathbb J}'_{lx}(\imath\hbar\vartheta) \rangle_{\rm eq} \, , \label{GK-L_kl}
\end{align}
where $\delta\hat{\mathbb J}'_{ab}=\delta\hat{\mathbb J}^{\prime a}_{g^b}$, $\delta\hat{\mathbb J}'_{e a}=\delta\hat{\mathbb J}^{\prime a}_{e}$, $\delta\hat{\mathbb J}'_{ka}=\delta\hat{\mathbb J}^{\prime a}_{n_k}$ for $a,b=x,y,z$, and $\delta\hat{\mathbb J}^{\prime}_{\zeta} \equiv ( \delta\hat{\mathbb J}^{\prime}_{xx}+\delta\hat{\mathbb J}^{\prime}_{yy}+\delta\hat{\mathbb J}^{\prime}_{zz})/3$ with the microscopic total currents~\eqref{micro-Je}-\eqref{micro-Jk}. The thermodiffusion and cross-diffusion coefficients obey the Onsager reciprocal relations ${\cal L}_{qk} = {\cal L}_{kq}$ and ${\cal L}_{kl} = {\cal L}_{lk}$.  These coefficients are not independent of each other, because the diffusive current densities should satisfy the constraints $\sum_{k} m_k {\cal J}^a_{n_k}=0$ as a consequence of mass conservation.  

Finally, the entropy production rate becomes
\begin{align}
\frac{dS}{dt} = & \int d{\bf r}\,  \frac{1}{T}\bigg[\, {\eta^{abcd} }\, \nabla^a v^b \, \nabla^c v^d 
 + \frac{\kappa}{T} \, \nabla^aT \, \nabla^a T \notag\\
&\qquad\qquad + 2 \sum_k \, \frac{{\cal L}_{qk}}{T} \, \nabla^a T \, \nabla^a\left(\frac{ \mu_{0k}}{T}\right) \notag\\
& + \sum_{k,l} {\cal L}_{kl} T \, \nabla^a\left(\frac{ \mu_{0k}}{T}\right)  \, \nabla^a\left(\frac{ \mu_{0l}}{T}\right)\bigg] \ge 0 \, ,\label{entropyproductionheatflow1}
\end{align}
where we have used the Onsager reciprocal relations between the transport coefficients.

\vspace{5mm}

\section{Phases with broken continuous symmetries}
\label{AppC}

The condensed matter phases with continuous symmetry breaking are characterized by microscopic local order fields $\hat{x}^{\alpha}$, which evolve in time according to
\be
\partial_t \, \hat{x}^\alpha + \hat{J}_{x^\alpha} = 0
\qquad\mbox{with}\qquad
 \hat{J}_{x^\alpha} \equiv \frac{1}{\imath\hbar} \, [\hat H, \hat{x}^\alpha]  \, .
 \ee
 Introducing the gradients of the order fields and the corresponding current densities according to
 \be
 \hat{u}^{a\alpha}\equiv \nabla^a \hat{x}^\alpha
 \qquad\mbox{and}\qquad
 \hat{J}^b_{u^{a\alpha}} \equiv \delta^{ab}\, \hat{J}_{x^\alpha} \, ,
 \ee
 these gradients obey the local conservation equations
 
 \begin{align}
 \partial_t \, \hat{u}^{a\alpha} + \nabla^b \hat{J}^b_{u^{a\alpha}} = 0 \, ,
 \end{align}
 which are comparable to the equations~\eqref{eqs-c} holding for the fundamentally conserved quantities.
In crystals, the order fields are given by the three Cartesian components of the displacement vector, which can be expressed in terms of the microscopic particle densities and the spatially periodic equilibrium densities of the different particle species in the crystalline lattice \cite{SE93,S97,MG20,MG21,H22,MGHF22}.

As in the classical case of Ref.~\cite{MG20}, the derivation of App.~\ref{AppB} can be extended to include the modes originating from the breaking of continuous symmetries. 


\end{document}